\documentclass[12 pt]{article}
\newcommand\ignore[1]{}
\renewcommand\footnote{\ignore}
\usepackage[left]{lineno}
\usepackage{amsmath}
\usepackage{graphicx}
\usepackage{braket}
\usepackage{authblk}
\usepackage{caption,subcaption}
\usepackage{comment}
\usepackage{enumitem}
\usepackage{tikz}
\usepackage{wasysym}
\usepackage{color}
\graphicspath{{fig/}}

\usepackage{amsmath}
\usepackage{amssymb}
\usepackage{graphicx}
\usepackage[margin = .75 in]{geometry}
\usepackage[pdftex, pdfstartview={FitH}\, pdfnewwindow=true, colorlinks=false, pdfpagemode=UseNone]{hyperref}
\newcommand\dd{\partial}
\newcommand{\nn}{\nonumber \\}
\newcommand\be{\begin{equation}}
\newcommand\ee{\end{equation}}
\newcommand\bea{\begin{eqnarray}}
\newcommand\eea{\end{eqnarray}}
\newcommand{\<}{\langle}
\renewcommand{\>}{\rangle}
\newcommand\half{ \textstyle {\frac{1}{2}}}
\newcommand\mR{ \mathbb R}
\newcommand\mT{ \mathbb T}

\newcommand\mRP[1]{ {\mathbb{RP}} (#1)}
\newcommand\mCP[1]{\mathbb{CP}(#1)}
\usepackage{placeins} 

\begin{document}

\thispagestyle{empty}

\begin{center}

{\Large \textsc{ Ising Model on the Affine Plane}}

\vskip15mm

Richard C. Brower$^{1}$,  Evan K. Owen$^1$

\vskip5mm


\it{$^1$Department of Physics, Boston University, Boston, MA 02215-2521, USA}\

\vskip5mm

\tt{brower@bu.edu, ekowen@bu.edu}

\end{center}


\begin{abstract}
  We demonstrate that the Ising model on a general triangular graph
  with 3 distinct couplings $K_1,K_2,K_3$ corresponds to an affine
  transformed conformal field theory (CFT).  Full conformal
  invariance of the $c= 1/2$ minimal CFT  is  restored by introducing a metric on the
  lattice through the map $\sinh(2K_i) = \ell^*_i/ \ell_i$ which relates critical
  couplings to the ratio of the dual hexagonal and triangular edge
  lengths.  Applied to a 2d toroidal lattice,  this provides an exact
  lattice formulation in the continuum limit to the Ising CFT as a function
  of the modular parameter.  This example can be viewed as a
  quantum generalization of the finite element method (FEM) applied to the
  strong coupling CFT at a Wilson-Fisher IR fixed point and suggests a
  new approach to conformal field theory on curved manifolds based on
  a synthesis of simplicial geometry and projective geometry on the
  tangent planes.
\end{abstract}

\newpage

\thispagestyle{empty}
\setcounter{page}{0}
\pagestyle{plain}
\setcounter{tocdepth}{2}
\setlength{\parskip}{.07in}
\tableofcontents
\setlength{\parskip}{.2in}
\newpage

\section{Introduction}

Ever since Onsager's famous solution in 1944 \cite{PhysRev.65.117},  
the 2d Ising lattice model has continued to stimulate a deeper
understanding of critical phenomena and conformal field theory
(CFT). On both a uniform square and triangular lattice, the 2d Ising model is equivalent to the
$c = 1/2$ minimal CFT on $\mR^2$ at the second order phase point. The
discrete translations  and  the  4-fold and 6-fold discrete
rotations (for the square and triangular lattice, respectively) are sufficient to guarantee the restoration of
Poincar\'{e} invariance (1 rotation and 2 translations).    Combined
with scale invariance at a second order fixed point, this apparently implies   
full conformal symmetry in the continuum limit~\cite{https://doi.org/10.48550/arxiv.2112.04811}.

\begin{figure}[h]
\begin{center}
\begin{tikzpicture}[>=stealth,scale=1.5]
    
    \foreach \y in {0,1,2,3}
    \foreach \x in {0,1,2,3} {
    \begin{scope}[shift={(\x+\y*0.5,\y*0.866)}]
    \draw[thick] (0,0) -- (1,0) -- (0.5,0.866) -- cycle;
    \draw[densely dashed] (0.25,0.433) -- (0.5,0.289) -- (0.5,0);
    \draw[densely dashed] (1,0.866) -- (1,0.866-0.289) -- (1.25,0.433);
    \draw[densely dashed] (0.5,0.289) -- (1,0.866-0.289);
    \draw (0.5,0.289) circle[radius=0.075];
    \draw (1,0.866-0.289) circle[radius=0.075];
    \end{scope}
    }
    \draw[thick] (2,4*0.866) -- (6,4*0.866) -- (4,0);

    \foreach \y in {0,1,2,3,4}
    \foreach \x in {0,1,2,3,4} {
    \begin{scope}[shift={(\x+\y*0.5,\y*0.866)}]
    \fill (0,0) circle[radius=0.075];
    \end{scope}
    }
    
    \begin{scope}[shift={(-1,2)}]
    \draw[thick,->] (0,0) -- (1,0);
    \draw[thick,->] (0,0) -- (-0.5,-0.866);
    \draw[thick,->] (0,0) -- (-0.5,0.866);
    
    \node[anchor=south] at (0.65,0.0) {$\hat{1}$};
    \node[anchor=south west] at (-0.25,0.5*0.866) {$\hat{2}$};
    \node[anchor=north west] at (-0.25,-0.5*0.866) {$\hat{3}$};
    \end{scope}
    
\end{tikzpicture}
\begin{tikzpicture}[scale=1.5]


\draw[thick] (-0.433,0) -- (1.732+0.433,0);
\draw[thick] (0.433,1.5) -- (2.598+0.433,1.5);
\draw[thick] (-0.433/2,-0.375) -- (0.866*1.25,1.5+0.375);
\draw[thick] (1.732+0.433/2,-0.375) -- (0.866*0.75,1.5+0.375);
\draw[thick] (1.732-0.433/2,-0.375) -- (2.598+0.433/2,1.5+0.375);
\draw[thick] (2.598-0.433/2,1.5+0.375) -- (2.598+0.433/2,1.5-0.375);
\draw[thick] (0.433/2,-0.375) -- (-0.433/2,0.375);

\fill (0,0) circle[radius=0.0833];
\fill (0.866,1.5) circle[radius=0.0833];
\fill (1.732,0) circle[radius=0.0833];
\fill (2.598,1.5) circle[radius=0.0833];

\node[anchor=south] at (0.866,0.0) {$K_1$};
\node[anchor=south] at (1.732,1.5) {$K_1$};
\node[anchor=north west] at (0.4,0.9) {$K_3$};
\node[anchor=north west] at (2.1,0.9) {$K_3$};
\node[anchor=south west] at (1.25,0.65) {$K_2$};

\end{tikzpicture}
\caption{\label{fig:graphsDuality} On the left, the triangular graph (black dots) and its hexagonal dual graph (open circles). On the right, the 3 independent couplings assigned
  to each triangle.}
\end{center}
\end{figure}
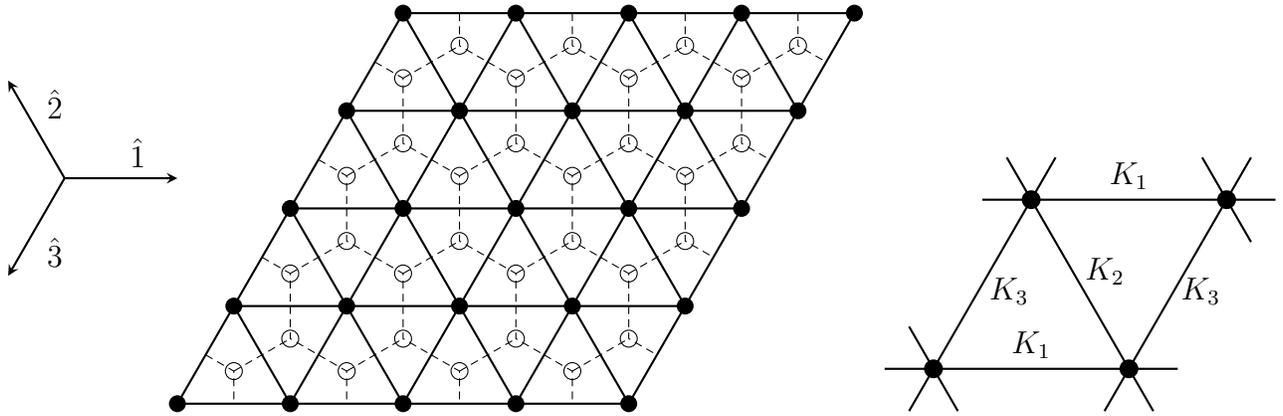

Here we consider the more general triangular Ising lattice  model partition function,
\be
Z = \sum_{s_n  =  \pm 1} e^{ \; \textstyle  K_1 s_n s_{n+\hat 1} +  K_2 s_n s_{n+\hat 2} +  K_3 s_n s_{n+\hat 3}   } \; ,
\label{eq:HexPartition}
\ee
with 3 independent couplings  ($K_1, K_2, K_3$) on the links connecting black sites illustrated in Fig.~\ref{fig:graphsDuality}.

Now, the map from the lattice to the continuum at a second order phase transition is  less obvious.
Indeed, even at  the second order
phase  transition, assigning equal lengths to the edges in the graph does {\bf not} in general restore
rotational  symmetry.  Although the action (\ref{eq:HexPartition})  has discrete  translations on three axes,  rotational symmetry   restricted to $\pi$
and reflections is insufficient to restore full rotational symmetry in the continuum limit except in the special cases
of the square  ($ K_1=  K_2, K_3= 0$) and equilateral triangular  ($K_1=  K_2= K_3$)  lattice.  Assuming a uniform lattice on  the edges of the triangular graph, in general the
geometry that emerges at the critical surface is instead an affine transformation of $\mR^2$. Affine transformations preserve
parallel lines but not angles, and map circles into ellipses. Consistent with these properties, the correlation length
contours of an affine-transformed CFT form ellipses. However, in the spirit of the finite element method (FEM), an alternative is to
introduce a metric on the graph by assigning unequal lengths
$\ell_1, \ell_2, \ell_3$ in correspondence with the couplings $K_1,K_2,K_3$. We
prove that the identity
\be
\sinh( 2 K_1) = \ell^*_1/\ell_1 \quad ,\quad \sinh( 2 K_2) = \ell^*_2/\ell_2 \quad ,\quad \sinh( 2 K_3) = \ell^*_3/\ell_3
\label{eq:IsingMap}
\ee
on the critical surface restores rotational symmetry in
the continuum. The dual  lengths  $\ell^*_i$ are defined as the lengths of the corresponding links on the hexagonal dual lattice, with dual sites defined to be at the circumcenter of each triangle in the original lattice.  While it  is dangerous to claim
originality for the classic  Ising model, we are unaware of this result in
the literature.  This result is of course consistent
with the special cases of the  Ising CFT  on  a uniform  square lattice ($\beta_c
=  K_1=  K_2, K_3= 0$) and equilateral triangular lattice ($\beta_3  = K_i$)
and the  well known critical temperatures:   $\beta_c    = \ln(1 + \sqrt{2}) $  with $\ell^*_i/\ell_i = 1$ and $\beta_3 =  \ln(3)/4$  with $\ell^*_i/\ell_i = 1/\sqrt{3}$, respectively.

Interestingly, this strong coupling solution (\ref{eq:IsingMap}) resembles the classic
FEM solution based on the discrete exterior calculus (DEC) for
a free scalar on a simplicial lattice. Applied to 
this  triangular graph, with action
\be
S_{\text{free}} =\frac{1}{2} \sum_n [ K_1(\phi_n  - \phi_{n+\hat 1})^2 +  K_2 (\phi_n - \phi_{n+\hat 2} )^2 +  K_3 (\phi_n  - \phi_{n+\hat 3})^2 ] 
\ee
the DEC prescription to converge to the continuum Laplacian on $\mR^2$  must satisfy the  condition 
\be
 2 K_1 = \ell^*_1/\ell_1 \quad ,\quad 2 K_2 = \ell^*_2/\ell_2 \quad ,\quad 2  K_3 = \ell^*_3/\ell_3 \; .
\label{eq:freeMap}
\ee
Thus, the free scalar 2d CFT is remarkably similar to the strong coupling Ising solution.   We note that  the  FEM represents
a very general solution to  free conformal field theory (CFT) in any dimension
for scalar, fermion, and gauge fields on a properly designed simplicial complex  for any smooth Riemann manifold.  This is a consequence of the equivalence of pure gaussian field theory to linear partial differential equation or equation of motion.
Our so-called quantum finite element (QFE) project ~\cite{Brower:2016vsl,Brower:2018szu,Brower:2020jqj}
is an attempt to extend the simplicial lattice
methods to quantum field theory on general Riemann manifolds -- a  much
more difficult problem  even for UV-renormalizable theories due to UV divergence.

In both Eq.~\ref{eq:IsingMap} and Eq.~\ref{eq:freeMap}, it
is instructive to regard {\bf geometry as
an emergent property} at a second order fixed point.  In lattice field theory
on regular (hypercubic) graphs, this feature is often subsumed in
the analysis of relevant operators at weak coupling at a UV fixed point.
Nonetheless, the geometry of a manifold for the quantum field theory is
dictated by the dynamics of the fixed point. Not only is this
the central concern of  the lattice Quantum Finite Element (QFE) project~\cite{Brower:2016vsl,Brower:2018szu,Brower:2020jqj} to define quantum field theory  on  curved  Riemann manifolds, but the Regge Calculus~\cite{Regge1961GeneralRW,Asaduzzaman:2022kxz} approach to gravity also hopes
to have space-time geometry  emerge at  distances far from the Planck scale. 

The  focus of this paper is to derive the identity in Eq. \ref{eq:IsingMap} and to understand if it has implications in a wider context.  To this end
it  appears natural and  possibly useful in exploring generalizations
to other systems to note how this problem relates to
in the language of  projective geometry.
We begin with a review in  Sec.~\ref{sec:Motivation}  of 
the fundamental elements at the intersection of  projective geometry and FEM discrete
exterior algebra calculus. Since the subsequent algebra is self-contained
the reader may  prefer to skip this section at first. 

In Sec. \ref{sec:StarTriangle} we  summarize the star-triangle
identity that maps the  critical surface of the hexagonal graph
to the triangular graph, followed by Sec. \ref{sec:FermonSoln} where we determine
the geometry by matching to a loop expansion of the theory of a free Wilson-Majorana fermion.
Interestingly, the formalism depends crucially on the simpler
properties of the trivalent form of the hexagonal graph.
Finally, in Sec. \ref{sec:Modular} we
show that this general projective Ising model allows a direct
computation of the general finite torus as a function of the modular
parameter, which was anticipated by Nash and O'Conner
in Ref.~\cite{nash1999modular}.   In Sec. \ref{sec:MCR2} we give numerical
verification of our projective Ising model on $\mR^2$ as the limit
of a finite toroidal lattice and in Sec. \ref{sec:RadialQ} we illustrate the
advantage of the radial quantization configuration in
the cylindrical geometry.  We hope that numerical methods may be able to extend
this approach to other theories that are not amenable to
analytical methods.  The conclusion in Sec. \ref{sec:Conclusion} suggests how this might be  extended  to
CFT on spheres or cylinders for radial quantization in the QFE program.

\section{\label{sec:Motivation} Motivation and Geometric Background}

\begin{figure}[h]
\begin{center}
\includegraphics[width=0.3\columnwidth]{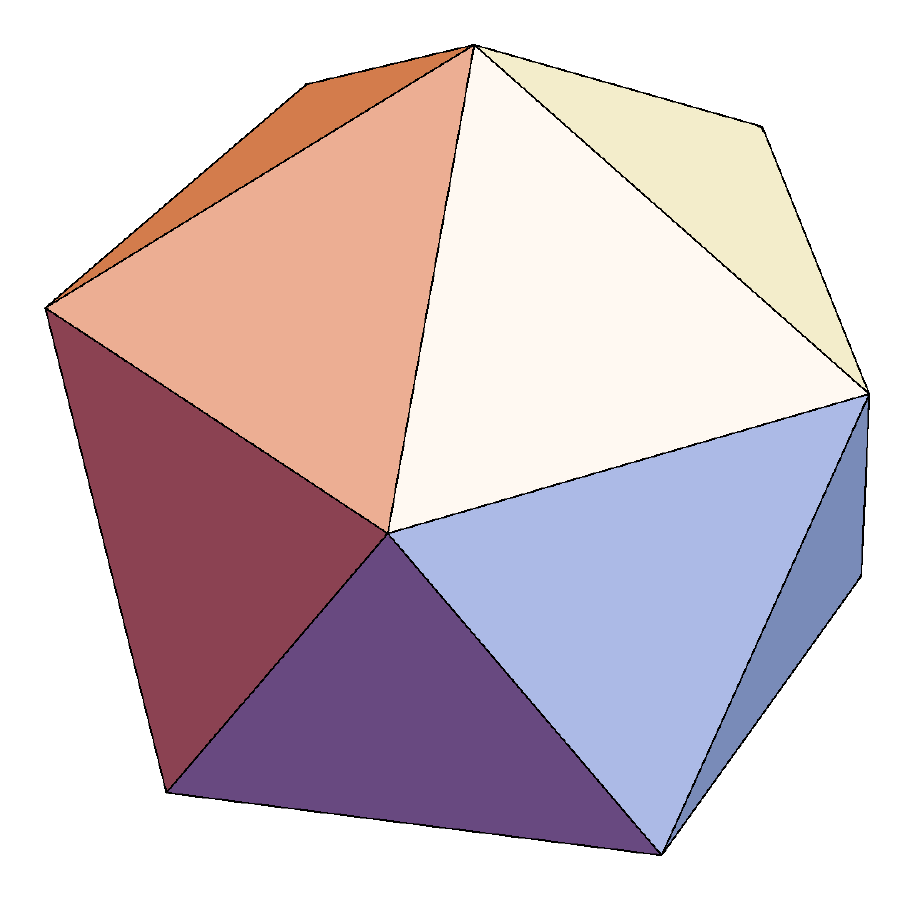}
\includegraphics[width=0.3\columnwidth]{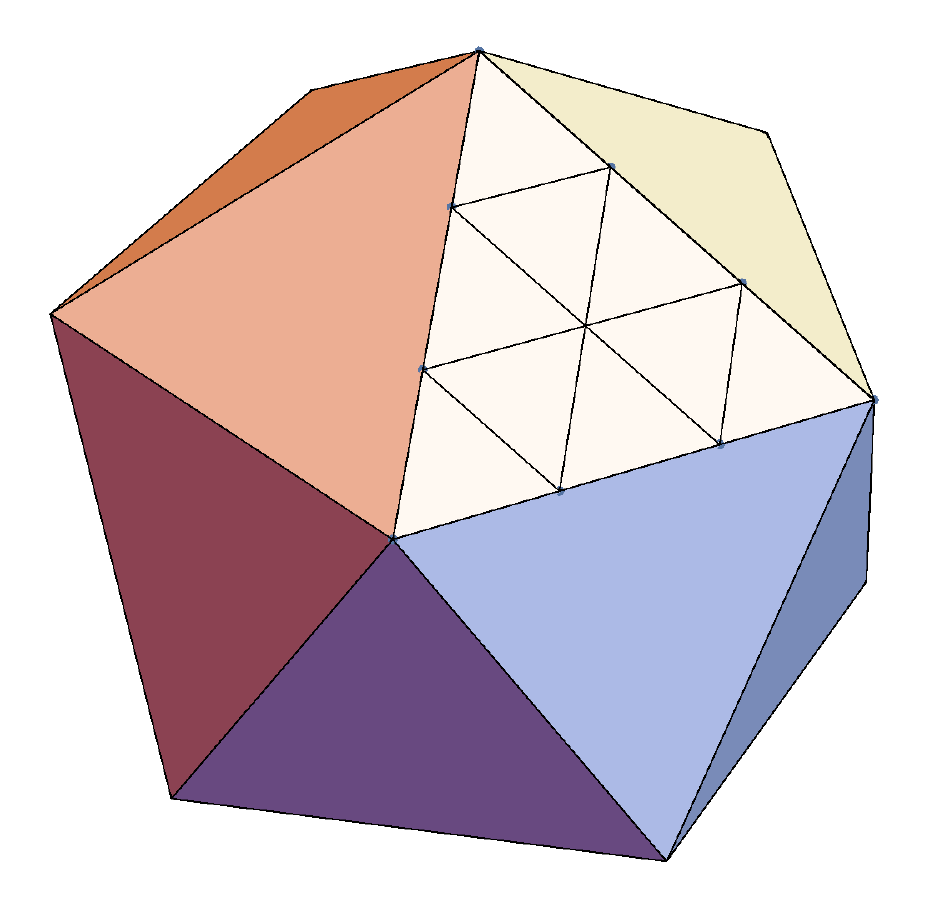}
\includegraphics[width=0.3\columnwidth]{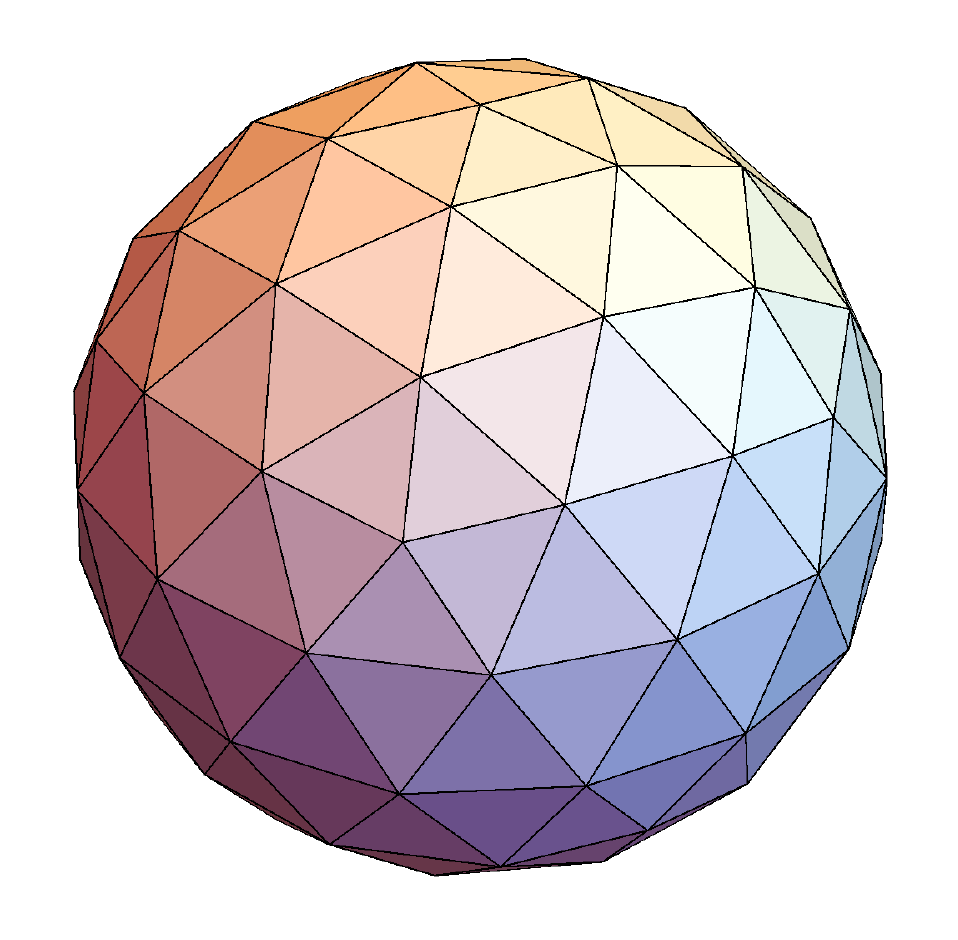}
\caption{\label{fig:icos} The $L$-th level of refinement of the icosahedron
subdivides triangles  into $L^2$ smaller triangles for a total of $N_\triangle = 20L^2$ faces,  $E = 30L^2$ edges and
$N = 2 + 10L^2$ sites.  Illustrated on the left the icosahedron, followed by the  $L=3$  refinement  with $ 2 + 10L^2 = 92$ vertices and finally
on the right subsequently projected onto the unit sphere.  }
\end{center}
\end{figure}

We first encountered the affine projection in the QFE project when
constructing a simplicial triangulation~\cite{Brower:2018szu} of
the sphere $S^2$. As illustrated in Fig.~\ref{fig:icos}, a  natural approach to a smooth simplicial triangulation was to
first refine the twenty flat faces of an icosahedron with equilateral triangles and then project each vertex radially outward onto a sphere of radius $R$, as illustrated in Fig.~\ref{fig:icos}. One advantage was that this  preserved exactly  the icosahedral subgroup of $O(3)$.

\begin{figure}[h]
\begin{center}
\includegraphics[width =0.75 \textwidth]{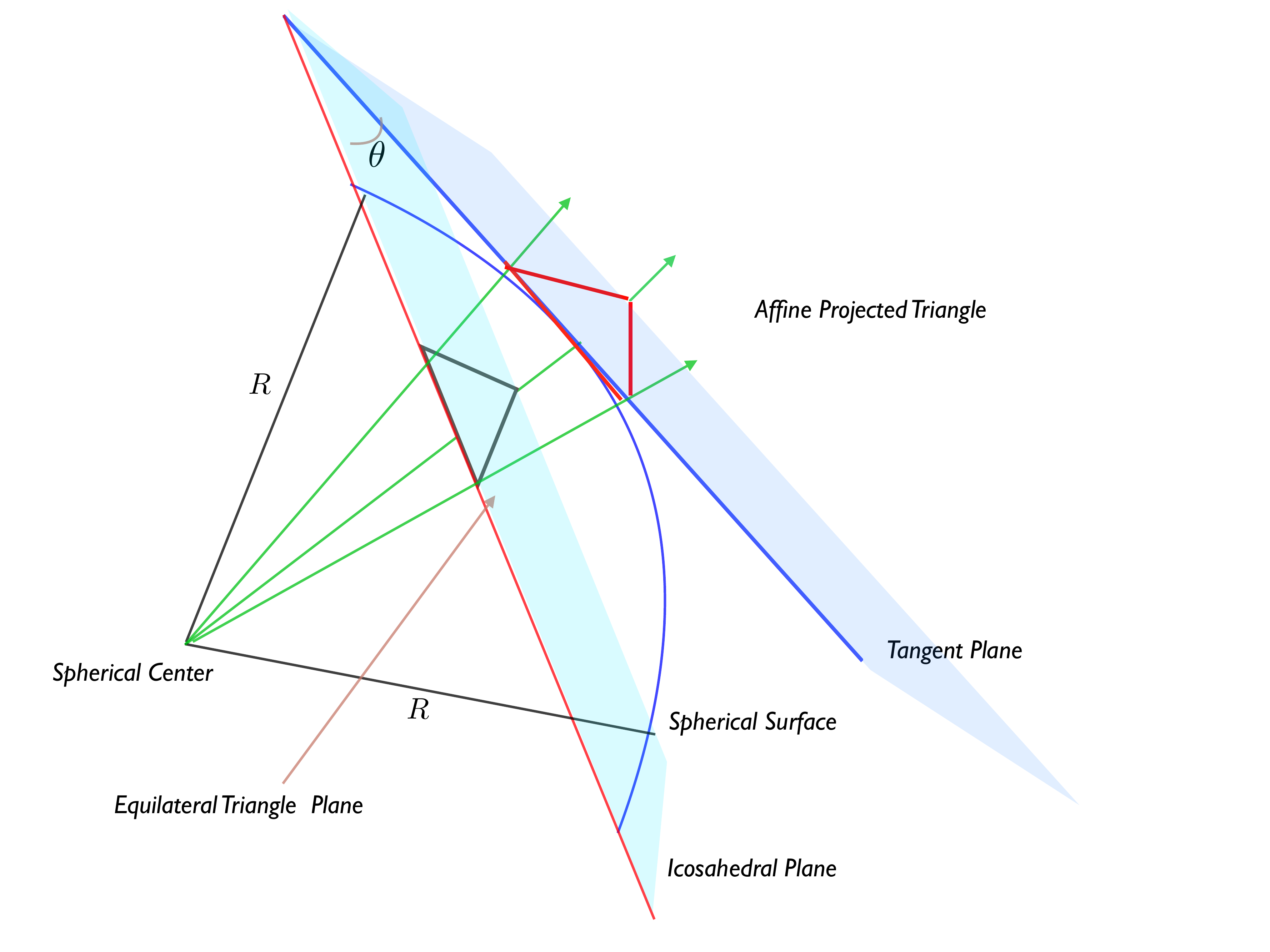}
\caption{\label{fig:AffineTan}  Projection of an equilateral triangular lattice to the tangent plane of a curved surface. Taking
the perspective point to infinity (or
equivalently taking the lattice spacing to zero), this projects the affine transformed triangular lattice
of each tangent plane to the flat space of our current investigation.}
\end{center}
\end{figure}

From the perspective point  at the center of the sphere, as illustrated in
Fig. \ref{fig:AffineTan}, in this construction each equilateral triangle undergoes an
affine transformation to the tangent plane. In the continuum limit of
infinitesimal triangles with edge length $\ell_i \sim a \rightarrow 0$
(or equivalently taking the perspective point to infinity) each
tangent plane converges to a uniform affine triangulation to $\mathcal{O}(a^2)$
introduced above in Eq.~\ref{eq:IsingMap}.
We are hopeful that the map to lattice couplings ($K_i$) will enable the determination of a
local quantum FEM action with a uniform UV cut-off of $\mathcal{O}(a^2)$ at each
point on the curved manifold.
The field of projective geometry is a vast and ancient discipline. 
While none of what we present in this section
is original or even essential to our algebraic construction, we think it helps guide the presentation
and to connect the topological structure
of the action on the left side of  Eq. \ref{eq:IsingMap}
to the metric information of the simplicial complex on the right side.

\subsection{Complex Projective Line} 
For 2d conformal field theory, the  stereographic projection
to the Riemann sphere~\cite{cite-key} adds a point at  complex infinity, which is a conventional model of the Complex Projective line $\mCP 1$.
In $\mCP 1$ a point is  defined by  homogeneous coordinates $(z^1,z^2) \sim \lambda (z^1,z^2) $  in the equivalence
class relative to complex scaling $\lambda$ and a line  $L_1,L_2$,  by a linear
homogeneous   form,  $L_1 z^1 + L_2z^2 = 0$. Thus, the automorphism of $\mCP 1$ is given
by a 2-by-2 complex matrix with unit determinant that take
lines into lines. Choosing a frame, $(z^1, z^2) = (z^1/z^2, 1) = (z,1)$, this
 transformation is
\be
\begin{bmatrix}
  a & b \\
  c & d 
\end{bmatrix}
\begin{bmatrix} 
 z  \\
 1  
\end{bmatrix}
=
\begin{bmatrix} 
 a z  + b \\
 c z + d  
\end{bmatrix}
\rightarrow
\begin{bmatrix}
  (a z  + b)/( c z + d) \\
  1
\end{bmatrix} \; . 
\ee
The linear algebra in homogeneous coordinates is 
equivalent to the M\"obius transform:  $\; z' = (a z  + b)/( c z + d)$ or the projection  $GL(2,C) \rightarrow PGL(2,C) $. A key feature of the M\"obius transformation on $\mR^2$ is that it preserve angles and takes
circles into circles with the line being a special circle of infinite radius. We note that now the point at infinity is
not special, so the Riemann sphere is an elegant manifold for studying 2d conformal theories.

\subsection{Real Projective Geometry and the Affine Plane}

On the other hand, $\mRP 2$ is a non-orientable
manifold represented by $S^2$ with antipodal points
identified.
In $\mRP d$, a point is represented by $d+1$ real numbers up to a real scale:
$X^I = (X^0,X^1, \cdots, X^d) \equiv \lambda X^I$. A line is defined by the scale-invariant homogeneous relation
$L_I X^I = 0$, or 
\be
L_1 x^1 + L_ 2 x^2  + L_3 x^3 = 0
\ee
in  $\mRP 2$. Projected
to a standard plane at $z = x^3 = 1$, this is a line $ a x + b y + c  = 0$ in the $x$-$y$ plane. The automorphism is a transformation with 8 real parameters,
\be
\begin{bmatrix}
  a_{11} & a_{12} & b_1 \\
 a_{21} & a_{22} & b_2\\     
  c_{1} & c_{2} & d
\end{bmatrix}
\begin{bmatrix}
 x^1 \\
  x^2 \\
 1
\end{bmatrix}
= 
\begin{bmatrix}
  a_{1j} x^j + b_1\\
  a_{2j} x^j + b_2\\
  {c_j x^j + d}
\end{bmatrix}
\rightarrow
\begin{bmatrix}
  (a_{1j} x^j + b_1)/(c_j x^j + d)\\
 (a_{2j} x^j + b_2)/(c_j x^j + d)\\
  1
\end{bmatrix} \; .
\ee
More interesting is  to contrast  the  affine 6-parameter group subgroup in $\mRP 2$ which leaves  the line at infinity fixed:
\be
 \vec x' = A \vec x + \vec d  \implies
\begin{bmatrix} 
 x' \\
 y'  \\
 1 
\end{bmatrix} = 
\begin{bmatrix}
  a_{xx} & a_{xy}  &d_x\\
  a_{yx} & a_{yy}  & d_y \\
  0 & 0 &1 
\end{bmatrix}
\begin{bmatrix} 
 x \\
 y \\
 1 
\end{bmatrix}
\ee
with the 6-parameter conformal M\"obius group. Both  groups extend the 4 parameters, the  Poincar\'{e} group  (2 translations, 1 rotation) and scaling but the 2 additional affine parameters  break conformal symmetry. They do not preserve angles or map
circles into circles.  Instead the additional affine transformations take circles into ellipses, with the additional parameters determining the eccentricity and shear direction of the major axis. All of these observations are generalized
in  comparing conformal vs. affine transformation in higher dimensions.

Finally, we can contrast this with the affine subgroup of the M\"obius group by fixing the complex point at infinity:
\be
\begin{bmatrix}
  a & b \\
  0 & 1 
\end{bmatrix}
\begin{bmatrix} 
 z  \\
 1  
\end{bmatrix}
=
\begin{bmatrix} 
 a z  + b \\
  1
\end{bmatrix} \; .
\ee
In real coordinates ($a = a_r + i a_i  = \lambda e^{i\theta}  , b = b_r, +   i b_i, z = x^1+ix^2$)
\be
z' = a z + b \implies 
\begin{bmatrix} 
x'  \\
y' \\
1
\end{bmatrix}
=
\begin{bmatrix}
  a_r & - a_i & b_r\\
  a_i &   a_r  &  b_i \\
   0 &  0 &  1 \\
 \end{bmatrix}
 \begin{bmatrix} 
x^1  \\
x^2\\
1
\end{bmatrix} \; .
\ee
In 2d, this is just the  4-parameter subgroup with Euclidean  Poincar\'{e} invariance
plus scaling common to both groups. 
\footnote{
 Geometry Algebra for  Vector spaces this
is sort of $
r = a^* r  + b =  a^* \wedge r + a^* \cdot  r +  b$ but I need to understand this better. There is  some text that does this
at \href{http://geocalc.clas.asu.edu/pdf/CompGeom-ch1.pdf}{http://geocalc.clas.asu.edu/pdf/CompGeom-ch1.pdf}} 

While beyond our current needs, a more general approach to conformal invariance is to
introduce two extra dimensions in $\mRP {d+1} $,  restricted to a null surface $- X^2_0 +  X^2_{d+1}  +   X^2_s + \cdots X^2_d = 0$,  invariant under the  Lorentz group  $O(d+1,1)$  and subsequently projected onto a light-like vector  $X_0 +X_{3} =1$ into $\mR^d$.  For example starting with $\mRP 3$
this induces the M\"obius transformation on $z = x_1 + i x_2$ identical to $\mCP 1 $ above.
Alternatively, adding two extra dimension and  fixing  $X^2_{d} = R^2 $,  the
hyperbolic geometry is the manifold for  Euclidean AdS$^{d+1}$ as in \cite{Brower:2022atv} with the $\text{CFT}^d$ residing at
the boundary.

\subsection{\label{sec:FreeCFT} Simplicial  Geometry and the Free CFT}

The affine transformation also plays a central role in simplicial geometry
and the finite element method (FEM)\footnote{See
\href{https://en.wikipedia.org/wiki/Barycentric\_coordinate\_system}{https://en.wikipedia.org/wiki/Barycentric\_coordinate\_system}   and \\
\href{https://en.wikipedia.org/wiki/Affine\_space\#Barycentric\_coordinates}{https://en.wikipedia.org/wiki/Affine\_space\#Barycentric\_coordinates. See Nima's generalization
of convex hull of vertices
over inside lines of a polytope. Can have n-points. Only invariant is epsilon symbol. Inside $A = p_1x_1 ... p_2 x_n$ \href{https://youtu.be/VdZSATkmeLs}https://youtu.be/VdZSATkmeLs} }. The FEM has a vast
literature with many implementations. Here is a glimpse of the
simplest and most geometrically intuitive methods based on a simplicial complex. 
A simplicial complex is  composed of  primitive $d$-simplices $\sigma_d(0,1,\cdots d)$
(corresponding to points, line segments, triangles, tetrahedrons, etc. for
$d = 0,1,2,3,\cdots$) with $d+1$ vertices and boundaries composed of $(d-1)$-simplices,  glued
together with the correct orientation required to define topology on the manifold. Adding edge  lengths 
fixes a  piecewise flat  metric interpolation of any  Riemann manifold. 

All of the primitive simplices are equivalent under affine transformations.
The $d(d+1)$ affine parameters match
the $d+1$ position vector $x^j$ in $d$-dimensions.  Primitive  simplices are  rigid objects. Exactly half of the $d(d+1)$ parameters define the simplex shape  determined completely by their $d(d+1)/2$ edge lengths. The other half are the Poincar\'{e}  rotations and translations to locate and orient the simplex. This decomposition is easily understood in terms of
the SVD form, $x = A x +b \equiv U \Sigma V + b$, 
where $U,V$ are rotations and the $d$ diagonal singular
values (or shearing  parameters) transform circles into ellipsoids.

The flat interpolation of the interior of each simplex  is parameterized
by  $d$ affine invariant barycentric coordinates $\xi_j$
\be
\label{eq:barycentric}
X(\xi_i) = \xi_0 x^0 + \xi_1 x^1 + \cdots + \xi_d x^d
\quad, \quad \xi_0 + \xi_1  + \cdots +  \xi_d = 1 \; .
\ee
 The discrete algebraic
properties of a simplicial complex and its Voronoi dual
are  remarkable, allowing the simplicial analogue of
exterior derivative, Hodge star, Stokes’ theorem etc. There is a
sequence of simplicial objects
$\sigma_0, \sigma_1, \cdots, \sigma_d$  (points, lines, triangles, etc.) and their Voronoi duals
$\sigma^*_d, \cdots \sigma^*_1, \sigma^*_0$, respectively, which are constructed recursively
from circumcenters. The circumcenters are found by the intersection of perpendicular lines
which gives the simplicial analogue of the $\epsilon^{i_1,i_2,\cdots, i_d} $ symbol central
to  both projective geometry and exterior calculus. 
For more details
the  reader is referred to Ref. \cite{Brower:2018szu} and the classical
papers for random lattices in flat space  by  Christ, Friedberg
and T.D. Lee~\cite{christ1982random,christ1982gauge,christ1982weights}.

\begin{figure}[h]
\begin{center}
  \includegraphics[width =0.48\textwidth]{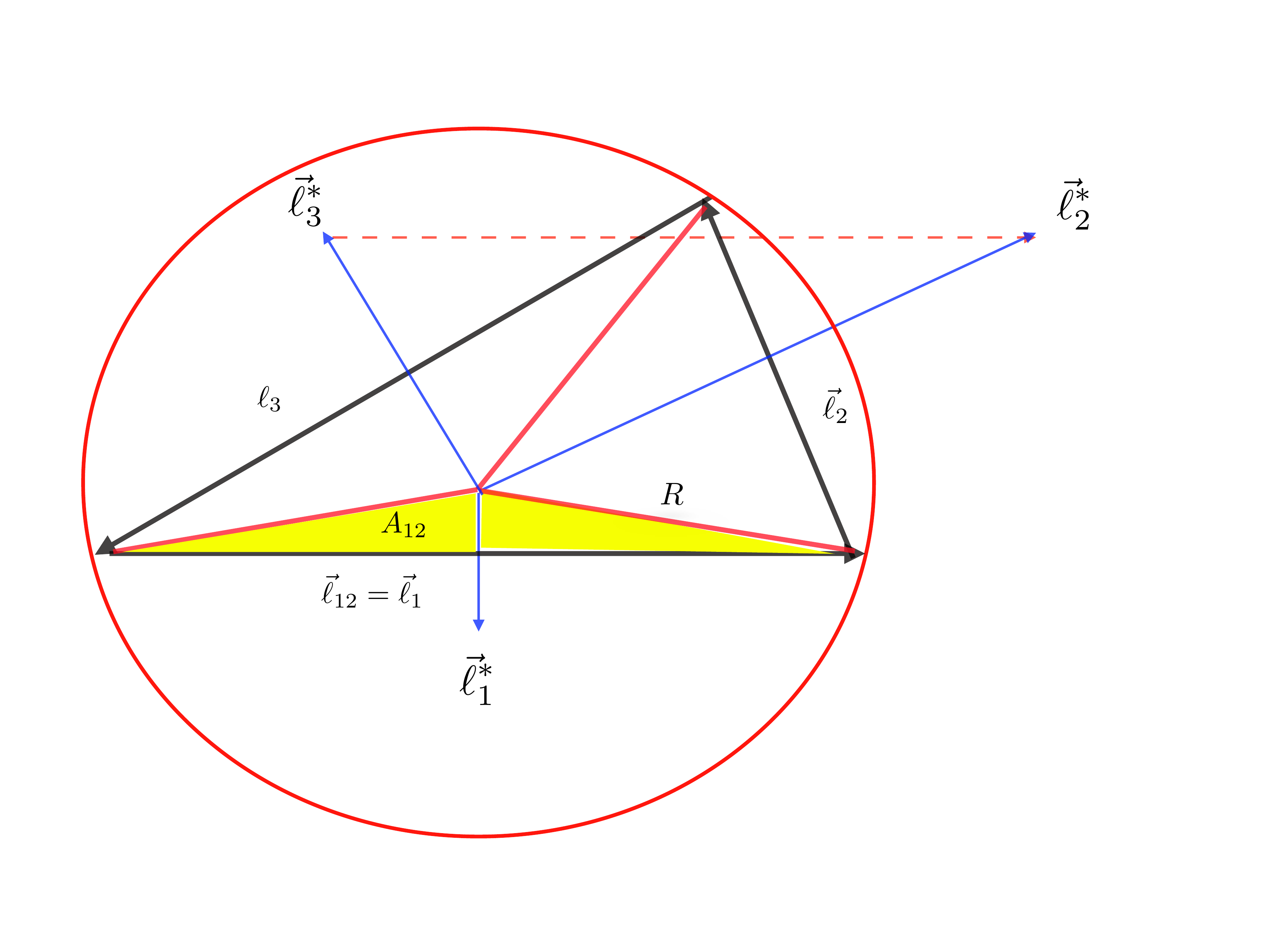}
   \includegraphics[width =0.5\textwidth]{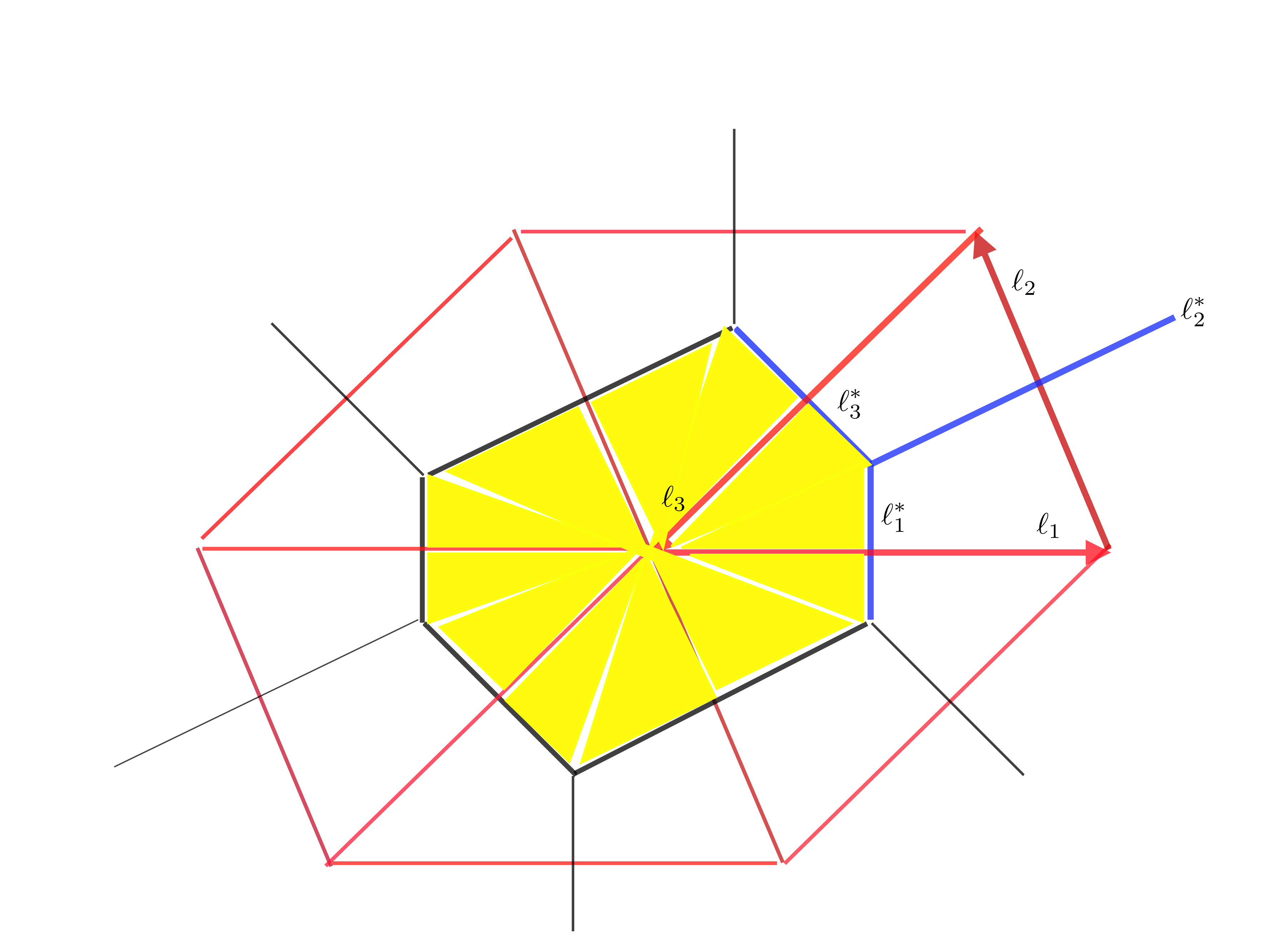}
\caption{\label{fig:Stokes} On the left is a general triangle defined by three vectors satisfying $\vec \ell_1 + \vec \ell_2+ \vec \ell_3 = 0$ and circumscribed
by the circle in red. In yellow is the
triangular area $A_{12}$ assigned to the free scalar FEM action. On the right is the discrete simplicial Beltrami-Laplace operator evaluated by flux for the
Voronoi hexagonal boundary in black for our affine lattice.}
\end{center}
\end{figure}

Applied to the continuum  free scalar  action,
\be
S =\half  \int_{\cal M} d^dx \sqrt{g} \left[ \;g^{\mu\nu}\dd_\mu
\phi(x)\dd_\nu \phi(x)\nn+ \left(m^2  + \frac{(d - 2)\text{Ric}}{4(d-1)} \; \phi^2(x) \right) \right] \; ,
\label{eq:scalar}
\ee
the  simplest FEM  approximation adds
a piecewise linear approximation (\ref{eq:barycentric}) to the manifold  and
a piecewise linear basis for the scalar fields in the interior
of each simplex (e.g. triangle),  $\phi(x) = \sum_i \xi_i \phi_i$. Performing the
integral in this basis gives
\be
S^{\text{(kin.)}}_\triangle = \frac{\ell^2_{23}+ \ell^2_{31} - \ell^2_{12}}{8 A_\triangle}(\phi_1 - \phi_2)^2 + (23) + (31) = \frac{1}{2}\frac{A_{12}}{ \ell^2_{12}}(\phi_1 - \phi_2)^2 + (23) + (31)
\label{eq:FEMtriangle}
\ee
for the kinetic term for the triangle  $\triangle(1,2,3)$ with  edge lengths $ \ell_{12},\ell_{23} ,\ell_{31} $ and  area $ A_\triangle$.  The intuitively
appealing form on the right side of Eq. \ref{eq:FEMtriangle} expresses the contribution to an edge from a single triangle as the area, $A_{12} = (\ell_{12}/2) \sqrt{R^2 + (\ell^2_{12}/2)}$ divided by the edge length squared, as illustrated on the left of Fig. \ref{fig:Stokes}. The organization of this geometry  in
any dimension on any manifold is the first job of the discrete
exterior calculus (DEC). In 2d, summing over any closed surface,  the scalar action is 
\be
S_{\text{FEM}} =  \frac{1}{2} \sum_{\<ij\>} \frac{\ell^*_{ij}}{ 2 \ell_{ij}} \; (\phi_i - \phi_j)^2
\label{eq:FEMaction}
\ee
where $\ell^*_{ij}$ is the distance between circumcenters of the adjacent
triangles sharing the edge $\<ij\>$.  In the language of
DEC, the equation of motion is the discrete simplicial Beltrami-Laplace operator 
\be
*d*d \phi_i =  *\frac{1}{|\sigma^*_0(i)|}\int_{\sigma^*_0} d[ *  (\phi_i -
\phi_j)/\ell_{ij}] 
=  \frac{1}{\sqrt{g_i}} \sum_{j \in \<i,j\>} \ell^*_{ij} \frac{\phi_i -\phi_j}{\ell_{ij}} \; .
\label{eq:BLoperator}
\ee
with a positive semi-definite spectrum and one null vector given by $\phi_i = \mbox{const}$. 
The discrete exterior derivative of a scalar is $d \phi_i = (\phi_j - \phi_i)/\ell_{ij}$  on the $\<i,j\>$ link and the integral
is a simplicial Stokes' theorem evaluated as the flux out of the dual Voronoi  polyhedron as illustrated in Fig. \ref{fig:Stokes}
for 2d.

In Fig. \ref{fig:Stokes} we have chosen to illustrate the  special 
case of our uniform  triangles which imposes the
identity, $\vec \ell_i = \epsilon^{ijk}(\vec \ell^*_j - \vec \ell^*_k)/2 $,
and the three FEM couplings in Eq. \ref{eq:IsingMap}:
\be
2K_1 = \ell^*_1/\ell_1 \quad ,\quad 2K_2 = \ell^*_2/\ell_2 \quad ,\quad 2 K_3 = \ell^*_3/\ell_3 \; .
\ee
These  projective  methods have a huge application domain, which
 includes a natural connection to simplicial geometry~\cite{Sobczyk_1992} and the discrete exterior calculus under
the rubric of conformal geometric algebra~\cite{https://doi.org/10.48550/arxiv.0907.5356}.

\section{\label{sec:StarTriangle} Star-Triangle Identity}

The star-triangle identity is a pure graph-theoretical result, mapping
between Ising spins on triangular and hexagonal graphs~\cite{Mussardo:2010mgq, HOUTAPPEL1950425}. There is no reference to distances or any metric.
To set conventions, consider the Ising model defined on a general
graph $G(N,E)$ with $N$ sites and $E$ undirected edges (or links)
$\<n,m\> \in G$ and partition
function,
\be 
Z^G_N[K] = \sum_{s_n =\pm 1} \exp \bigg[ \sum_{\<n,m\>} K_{nm} s_n s_m \bigg]
\equiv \operatorname{Tr}_s \bigg[ e^{\textstyle K_{nm} s_n s_m} \bigg]
\ee
where couplings with $K_{nm} = 0$ imply the absence of links in graph.

Specializing to a triangular graph and its dual hexagonal graph
illustrated in Fig. \ref{fig:graphsDuality}, each with three distinct
couplings $K_1,K_2,K_3$ and $L_1,L_2,L_3$, respectively, gives
\be 
Z^\triangle_N(K) = \operatorname{Tr}_s \bigg[ e^{\textstyle K_1 s_n s_{n + \hat 1} + K_2 s_n s_{n + \hat 2} + K_3
   s_n s_{n + \hat 3}} \bigg]
\ee 
and 
\be
Z^{\hexagon}_N(L)= \operatorname{Tr}_s \bigg[ e^{\textstyle \sum_{n/2} L_1 s_n s_{ n +\hat 1}+
 L_2 s_n s_{ n +\hat 2}+ L_3 s_n s_{ n +\hat 3} } \bigg] \; . 
\ee
The index $ n + \hat i$ for $i =1,2,3$
enumerates the distinct non-zero links $\<n,n + \hat i\>$. For
high temperature power series  expansion we introduce the
parameters  $v_i = \tanh(K_i)$ and
$t_i = \tanh(L_i) $. The star-triangle identity gives an exact map between the two
partition functions.

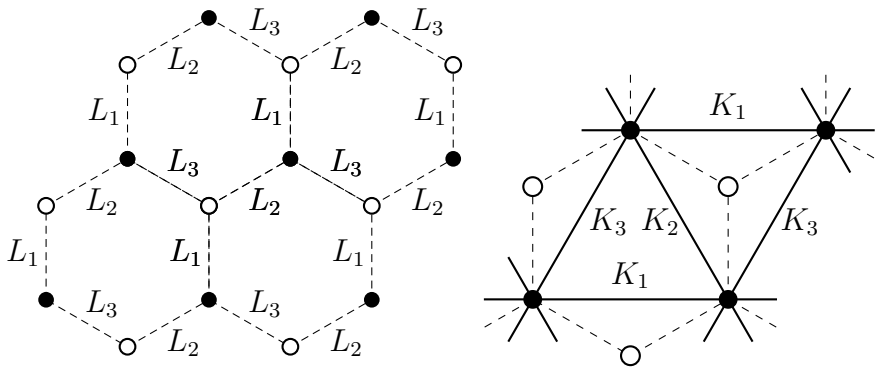
\begin{figure}[h]
\begin{center}
  \begin{tikzpicture}[scale=1.25]
    
\foreach \y in {0,1}
\foreach \x in {0,1} {
    \begin{scope}[shift={(\x*1.732+\y*0.866,\y*1.5)}]
    \draw[densely dashed] (0,0) -- (0,1) -- (0.866,1.5) -- (1.732,1) -- (1.732,0) -- (0.866,-0.5) -- cycle;
    
    \fill (0,0) circle[radius=0.0833];
    \fill (0.866,1.5) circle[radius=0.0833];
    \draw [thick,fill=white] (0,1) circle[radius=0.0833];
    \draw [thick,fill=white] (0.866,-0.5) circle[radius=0.0833];
    
    \node[anchor=east] at (0.05,0.5) {$L_1$};
    \node[anchor=east] at (0.05+1.732,0.5) {$L_1$};
    \node[anchor=south west] at (0.3,-0.3) {$L_3$};
    \node[anchor=south west] at (0.3+0.866,-0.3+1.5) {$L_3$};
    \node[anchor=north west] at (0.3,1.3) {$L_2$};
    \node[anchor=north west] at (0.3+0.866,1.3-1.5) {$L_2$};
    \end{scope}
}

\fill (4.33,1.5) circle[radius=0.0833];
\fill (3.464,0) circle[radius=0.0833];
\draw [thick,fill=white] (4.33,2.5) circle[radius=0.0833];

\end{tikzpicture}
  \begin{tikzpicture}[scale=1.5]

\draw[dashed] (0,0) -- (0,1) -- (0.866,1.5) -- (1.732,1) -- (1.732,0) -- (0.866,-0.5) -- cycle;
\draw[dashed] (0,0) -- (-0.433,-0.25);
\draw[dashed] (0.866,1.5) -- (0.866,2.0);
\draw[dashed] (1.732,1) -- (2.598,1.5);
\draw[dashed] (2.598,1.5) -- (2.598,2.0);
\draw[dashed] (2.598,1.5) -- (3.031,1.25);
\draw[dashed] (1.732,0) -- (2.165,-0.25);

\draw[thick] (-0.433,0) -- (1.732+0.433,0);
\draw[thick] (0.433,1.5) -- (2.598+0.433,1.5);
\draw[thick] (-0.433/2,-0.375) -- (0.866*1.25,1.5+0.375);
\draw[thick] (1.732+0.433/2,-0.375) -- (0.866*0.75,1.5+0.375);
\draw[thick] (1.732-0.433/2,-0.375) -- (2.598+0.433/2,1.5+0.375);
\draw[thick] (2.598-0.433/2,1.5+0.375) -- (2.598+0.433/2,1.5-0.375);
\draw[thick] (0.433/2,-0.375) -- (-0.433/2,0.375);
    
\fill (0,0) circle[radius=0.0833];
\draw [thick,fill=white] (0,1) circle[radius=0.0833];
\fill (0.866,1.5) circle[radius=0.0833];
\draw [thick,fill=white] (1.732,1) circle[radius=0.0833];
\fill (1.732,0) circle[radius=0.0833];
\draw [thick,fill=white] (0.866,-0.5) circle[radius=0.0833];
\fill (2.598,1.5) circle[radius=0.0833];

\node[anchor=south] at (0.866,0.0) {$K_1$};
\node[anchor=south] at (1.732,1.5) {$K_1$};
\node[anchor=north west] at (0.4,0.9) {$K_3$};
\node[anchor=north west] at (2.1,0.9) {$K_3$};
\node[anchor=north east] at (1.4,0.9) {$K_2$};

\end{tikzpicture}
  \caption{\label{fig:bipatite} On the left the bipartite structure of the hexagonal graph with its coupling parameters $L_1,L_2,L_3$  and on right
    the star-triangle  map to the triangular graph with coupling constants $K_1,K_2,K_3$.  }
\end{center}
\end{figure}

Starting with the hexagonal graph,  we note that it is bipartite (see  Fig. \ref{fig:bipatite}),
coupling black and white sites. 
Consequently we may \textit{decimate} the hexagonal graph by summing over
each  white spin and projecting it onto the triangular model. For example, suppose 
a specific
spin $s_0$ on a white vertex is connected to the neighboring black spin $s_1,s_2,s_3$ by edges with couplings $L_1,L_2,L_3$ as illustrated in Fig. \ref{fig:bipatite} on the right.
The result is the local star-triangle identity
\be
\sum_{s_0 = \pm 1} e^{\textstyle L_1 s_0 s_1 + L_2 s_0 s_2+ L_3s_0 s_3 } = {\cal D} e^{\textstyle K_1 s_2 s_3 + K_2 s_3 s_1+ K_3 s_1 s_2 }
\label{eq:StarTriangle}
\ee
with  ${\cal D}^2 = 2 h \sinh 2L_1 \sinh 2L_2\sinh 2L_3$. 
Obviously both the RHS and LHS side of Eq.~\ref{eq:StarTriangle} can be
expanded into four terms: $ c_0 + c_1 s_2 s_3 +  c_2 s_3 s_1 + c_3 s_1
s_2 $. Choosing four independent configurations $(s_1,s_2, s_3) = (1,1,1), (-1,1,1), (1,-1,1), (1,1,-1)$ determines
a  4 parameter map: $(L_1,L_2,L_3, {\cal D})$ $\leftrightarrow$ $(K_1K_2,K_3, h )$. The algebra is intricate and  is  probably  best
understood as special case of the soluble models and the Yang-Baxter
relation~\cite{Mussardo:2010mgq}, nonetheless the result is given by
\be
h(K_1,K_2,K_3) = \frac{ (1 - v_1^2) (1 - v_2^2) (1 - v_3^2)}{ 4 \sqrt{(1 +  v_1 v_2 v_3)(v_1 +  v_2 v_3) (v_2 +  v_3 v_1) (v_3 +  v_1 v_2)}}
\label{eq:Hparam}
\ee
with $v_i =  \tanh(K_i)$ and
\be
h \sinh( 2K_1) \sinh(2L_1) = h \sinh( 2K_2) \sinh(2L_2) = h \sinh( 2K_3) \sinh(2L_3) = 1 \; .
\ee
The consequence is an {\bf exact equivalence}  of the hexagonal and triangular
partition functions for all values of the couplings, 
\be
Z^{\hexagon}_{2N}(L_i)  = {\cal D}^N  Z^\triangle_N (K_i) = ( 2 h \sinh(2L_1)\sinh(2L_2)\sinh(2 L_3) )^{N/2}   Z^\triangle_N (K_i) \; .
\label{eq:Decimation} 
\ee
{\bf Note that this is a graph-theoretic result}  giving a one-to-one mapping:  $L_i \leftrightarrow K_i$. Although there is
no need for a reference to distances,  if the vertices can be placed at regular intervals in $\mR^2$, 
in the spirit of Wilsonian blocking both partition functions
will share a second order critical point. If such a point exists, the theories will have identical correlation functions in the infrared. 

The next step  is to find the critical surface, assuming there is a single transition, which is well known in the limit of the triangular ($K_1 = K_2= K_3 $) and
square ($K_1 = K_2, K_3 = 0$) lattice
Ising  models. The procedure is a generalization of the Kramers-Wannier
duality \cite{Kramers},  comparing  the weak coupling (high temperature) 
expansion on the triangular Ising graph with the strong coupling (low temperature) expansion on the hexagonal graph.
At high temperature, expanding in $v_i = \tanh(K_i) $  on the triangular graph, we have
\be
Z^\triangle_N(K) = ( \cosh K_1 \cosh K_2  \cosh K_3 )^{N} \operatorname{Tr}_s \bigg[ \prod_{\<n,i\>}  \left( 1 + v_i s_n s_{n + \hat i} \right) \bigg] \; .
\ee
The notation $\<n,i\>$ is used to specify the undirected link $\<n,m\> = \<n, m = n + \hat i\>$.
Summing over  the spins, we must pair spins on each site,
giving  all paths ${\cal P}$ with a product of links weighted by $v_i$.

Now compare with  the hexagonal graph, expanding at low temperature in powers of  $q_i  = e^{ - 2 L_i}$ ,
\be
Z^{\hexagon}_{2N}(L_i)  =  e^{ N (L_1 + L_2 + L_3)}  \operatorname{Tr}_s \bigg[\prod_{n,i} \left( q_i   + (1 - q_i)\delta_{s_n, s_{n + \hat i}} \right) \bigg]  \; ,
\ee
where we have used the identity:  $e^{L_i  s_n s_{n + \hat i}}   =   e^{L_i}[ q_i   + (1 - q_i) \delta_{s_n, s_{n + \hat i}}]$. The sum over spins enumerates the number of  broken (anti-parallel) bonds  entering and exiting each hexagon on the  dual
triangular lattice. Since the $N$-site  triangular graph is dual to the $2N$-site hexagonal graph, the 
loop expansions are identical  at $ e^{ -2 L_i}  \equiv\tanh (K_i) $ proving,
\be
Z^{\hexagon}_{2N}(L_i)  =  (2  \sinh2L_1 \sinh 2L_2 \sinh 2L_3  )^{N/2}   Z^\triangle_N (K_i)
\label{eq:KWduality} \; .
\ee
Convergence is subtle but we assume a proper limit to the critical surface.
Combining the star-triangle identity (\ref{eq:Decimation}) and the duality condition (\ref{eq:KWduality})
proves that if there is a single common phase boundary for both the hexagonal and triangular partition functions,  the critical
surface satisfies $h(K_1,K_2,K_3) = 1$ and the duality map $\sinh 2K_i \sinh 2L_i =1$.
Introducing the  pair of high/low expansion coefficients
\be
v_i = \tanh(K_i) \; , \; p_i = e^{ - 2K_i}  \quad \mbox{and }  \quad t_i = \tanh(L_i)\; , \; q_i = e^{ - 2L_i}
\ee
for the triangular  and hexagonal models respectively this is equivalent to
the duality condition $p_i = t_i$ or  $q_i = v_i$.

Setting $h =1$ in  Eq.~\ref{eq:Hparam} results in  a 12th order  polynomial~\cite{Mussardo:2010mgq}, in  $v_i = \tanh(K_i)$ but we find that
this is can be reduced to an elegant form ~\footnote{ This result from a factorization of the polynomial the 12 order polynomial  is factorized into four 3rd order terms with one permutation invariant factor, $ 1 - v_1 - v_2 - v_3 - v_1 v_2 - v_2 v_3 - v_3 v_1   + v_1 v_2 v_3$,
which is using $v_i = (1 - p^2_i)/(1 + p^2_i)$ vanishing on the critical surface and expressed as equivalent to  the quadratic, $p_1 p_2 +   p_2 p_3 + p_3 p_1 =1$. The are 3  other factors where a pair $v_i, v_j $  are at unphysical couplings.}
\be
p_1 p_2 + p_2 p_3 + p_3 p_1 = 1  \quad \mbox{with} \quad p_i = \exp( - 2 K_i)
\label{eq:CriticalSurface}
\ee
for the physically relevant roots. Equivalently, for the hexagonal lattice the surface is 
\be
t_1 t_2 + t_2 t_3 + t_3 t_1 = 1  \quad \mbox{with} \quad t_i =  \tanh(L_i)  \; .
\ee
This has several interesting properties.  These probability
weights ($ 0 < p_i \le 1$)   enter the Swendsen-Wang cluster  algorithm to cut bonds between aligned spins \cite{SwendsenWang}.   Likewise  $q_i = \exp( - 2 L_i)$ are dual Swendsen-Wang   probabilities on the  hexagonal lattice.  They obey  a  symmetric
M\"obius  map $p_i =  (1 -q_i)/(1+q_i)$ and  $q_i =  (1 -p_i)/(1+p_i)$.

The question we raise in Sec. \ref{sec:FermonSoln} is whether the dynamics induces a metric for correlation functions. More specifically,
\begin{itemize}
\item For the critical theory, is there a lattice metric (e.g. a set of edge lengths)
that restores  rotational symmetry on $\mR^2$ on the critical surface?
\item Is this map on the finite torus consistent with
modular invariance for the continuum $c =1/2$ minimal CFT?
\end{itemize}
Both of these we answer in the affirmative and support by  numerical evidence.

\section{ \label{sec:FermonSoln} Wilson-Majorana Fermionic Solution}

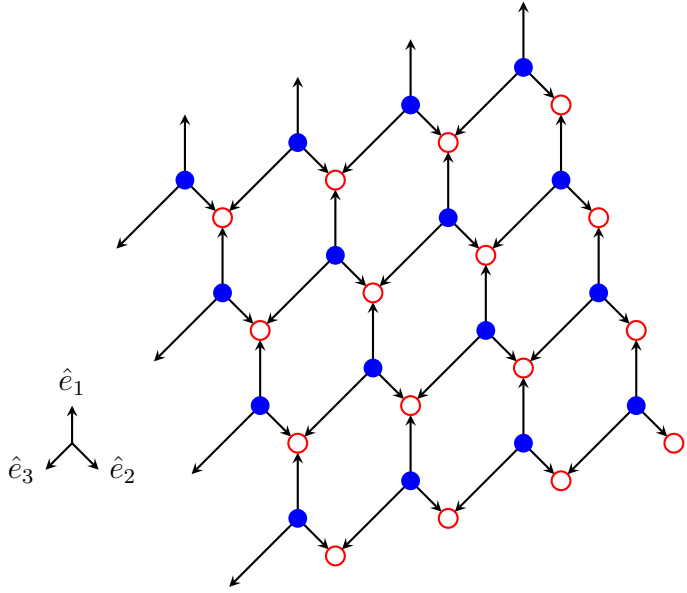
\begin{figure}[h]
\begin{center}
\begin{tikzpicture}[scale=0.5,>=stealth]
    
\foreach \y in {0,1,2,3}
\foreach \x in {0,1,2,3} {
\begin{scope}[shift={(\x*3+\y*-1,\x*1+\y*3)}]
    \draw[->,thick,shorten >=0.125cm] (0,0) -- (0,2);
    \draw[->,thick,shorten >=0.125cm] (0,0) -- (1,-1);
    \draw[->,thick,shorten >=0.125cm] (0,0) -- (-2,-2);

    \fill[blue] (0,0) circle[radius=0.25cm];
    \draw[red,thick,fill=white] (1,-1) circle[radius=0.25cm];
    
    \end{scope}
}

\begin{scope}[shift={(-6.0cm,2cm)}]

    \draw[->,thick] (0,0) -- (0,1);
    \draw[->,thick] (0,0) -- (0.707,-0.707);
    \draw[->,thick] (0,0) -- (-0.707,-0.707);
    
    \node[anchor=south] at (0,1) {$\hat{e}_1$};
    \node[anchor=west] at (0.707,-0.707) {$\hat{e}_2$};
    \node[anchor=east] at (-0.707,-0.707) {$\hat{e}_3$};
\end{scope}

\end{tikzpicture}
\caption{\label{fig:hgeo} Hexagonal lattice used in our derivation showing the bipartite and trivalent structure. 
}
\end{center}
\end{figure}

In order to find the induced metric at the critical point (\ref{eq:CriticalSurface}),
we use a  special property of the 2d Ising model. The partition
function can be represented on the graph as a
free   Wilson-Majorana fermion~\cite{kac1952combinatorial}, which is a
free CFT at the critical point and fixes the emergent
geometry by FEM methods~\cite{Brower:2016vsl}.
Our approach follows Wolff's elegant  paper~\cite{Wolff:2020oky} on
the fermionic representation of the  uniform hexagonal lattice generalized
to the  3 parameter coupling space: $L_1,L_2,L_3$.  Wolff notes that the
loop expansion on the hexagonal lattice is simpler than that of the square or triangular lattice because there are no self-intersections due to the trivalent form shown in
Fig. \ref{fig:hgeo}.
Indeed this argument can be  extended to any
trivalent dual of a triangular simplicial graph, but for simplicity we restrict the
analysis to our three parameter hexagonal graph.

The high temperature expansion in $t_i  = \tanh L_i $
is 
\be
Z^{\hexagon}_N(L_i)  = \operatorname{Tr}_s \bigg[ e^{\textstyle \sum_{\<n,i\>} L_i \, s_{n} s_{n +\hat i}} \bigg] = (\cosh L_1 \cosh L_2 \cosh L_2)^{N/2}\; \operatorname{Tr}_s \bigg[ \prod_{\<n,i\>} (1 + t_i s_ n s_{n + \hat i}) \bigg]
\label{eq:Zhex} \; .
\ee
Again the notation $\<n,i\>$ implies a product over each undirected link, or
equivalently the restriction of $n$ to blue sites only. Summing over the spins leads to a loop expansion of all non-self intersecting closed  paths
${\cal P}$ of length $\Lambda = n_1 + n_2 + n_3$ with power $t_1^{n_1} t_2^{n_2} t_3^{n_3} $.

We now introduce a hexagonal lattice action for Wilson-Majorana Fermions, 
\be
S[\psi] = \dfrac{1}{2} \sum_n \bar{\psi}_n \psi_n -  \sum_{n, i}  \kappa_i \bar{\psi}_{n} P (\hat{e}_i) \psi_{ n + \hat i} 
\label{Smajo}
\ee
and demonstrate that its partition function 
\be
Z^\psi_N = \prod_n  \iint d\psi^1_n d\psi^2_n \; e^{ \textstyle - S[\bar \psi, \psi]}  = 
\prod_n  \int d^2 \psi_n e^{\textstyle  -\frac{1}{2} \sum_n \bar \psi_n \psi_n}  \prod_{n,i}
\left[ 1 +  \kappa_i \bar{\psi}_{n} P (\hat{e}_i) \psi_{ n + \hat i} \right]
\label{eq:DiracPartion}
\ee
expanded in $\kappa_i$ matches the Ising loop expansion (\ref{eq:Zhex}) term by term. For the sums and products of nearest-neighbor spinors above, it is understood that each directed link is included only once.
The Wilson link matrices are
\be
P (\hat{e}_i) = \frac{1}{2}( 1 + \hat e_i \cdot \vec \sigma)  \quad, \quad \vec \sigma = (\sigma_1 ,\sigma_2)  \; .
\ee
{\bf The  introduction of unit vectors $\hat e_i$ is intentional.}  In general, Fermionic spinors
on a simplicial complex   (e.g a triangulated lattice) require tangent vectors
$\hat e_i$   for the lattice vierbein $E_i = \hat e_i \cdot \vec \sigma$.
Because we anticipate a critical theory in flat space, the tangent vectors on all 
planes can be given
in global coordinates in a gauge with
zero spin connection. 

In order for the Grassmann integral in Eq.~\ref{eq:DiracPartion} to match the non-self intersecting
loop expansion, there must be only two fermionic variables per site. 
This is accomplished by replacing
4 Grassmann Dirac variables  $\bar \psi, \psi$ 
by 2  Wilson-Majorana  fermions~\cite{Wolff:2020oky}  
that obey the charge conjugation constraint, $\bar
\psi = \psi^T {\cal C}$: 
\be
{\cal C} = - i \sigma_2 =
\begin{bmatrix}
0 & 1\\
- 1 & 0
\end{bmatrix} 
\quad, \quad \bar \psi^1 =  - \psi^2 \quad \mbox{and} \quad \bar \psi^2 = \psi^1
\ee
or $\bar \psi^j = \psi^i\epsilon_{ij} $
with the epsilon operator,  ${\cal C}_{ij} = \epsilon_{ij}$. The mass term is
\be
e^{\textstyle  -\frac{1}{2} \sum_n \bar \psi_n \psi_n}  = \prod_n \left[ 1 - \frac{1}{2} \psi^i_n \epsilon_{ij}\psi^j_n \right]
= \prod_n \left[1 - \frac{1}{2}(\psi^1_n \psi^2_n  - \psi^2_n \psi^1_n) \right]
\ee
and the Grassmann measure at each site is defined as $d^2\psi_n =
d\psi^1_n d\psi^2_n$.

To generate the loop expansion, we compute
the product of Wilson matrices. In flat 2d space with no-self intersections,
the winding number is $\pm 1$ resulting in a spinor  phase $\exp[ i \theta/2]= -1 $ with $\theta =  \pm 2 \pi$. The  minus sign is cancelled by the anti-commutation of the Grassmann measure over
the closed path~\footnote{Careful order counts:
  $\int d\psi^1 d\psi^2 \psi^2 \bar \psi^1 = - \int d\psi^1 d\psi^2 \psi^1
  \bar \psi_2 = \int d\psi_1 (\int d\psi_2 \bar \psi_2)\psi_1 = 1 $
  using $\psi^1 \psi^2 = -
  \psi^1 \psi^2 $ or $d\psi^1 d\psi^2 = - d\psi^2 d\psi^1$.
  Grassmann integration is equivalent to ordered derivatives:
  $\int d\psi \; 1 = 0$ and $\int d\psi \psi = 1$. There are no higher
order terms because $\psi^2 = 0$ so $\int d\psi f(\psi) = f'(0)$! }. To make this concrete we will carry this out explicitly.
In global coordinates, defining $\hat e_i = (\cos\theta_i,\sin \theta_i)$, the  Wilson factor is a (dyadic) projection matrix:
\be
P (\hat{e}_i) = \frac{1}{2}( 1 + \hat e_i \sigma)  = \frac{1}{2}
\begin{bmatrix}
1 &   e^{- i \theta_i}\\
e^{ i \theta_i} &1 
\end{bmatrix}
= \frac{1}{2}
\begin{bmatrix}
1 \\
e^{i \theta_i}  
\end{bmatrix}
\begin{bmatrix}
1 & e^{- i \theta_i}  
\end{bmatrix} \; .
\ee
This gives a spinor  rotation phase at each  corner,
\be
R_{z}(\theta_{21}) = \frac{1}{2}
\begin{bmatrix}
1 & e^{- i \theta_1}  
\end{bmatrix}
\begin{bmatrix}
1 \\
 e^{ i \theta_2}  
\end{bmatrix}
= \cos(\theta_{12}/2) e^{ i (\theta_2 - \theta_1)/2}
\ee
by $(\theta_2 - \theta_1)/2$ at each vertex giving rise
to a minus sign (or by $e^{i \pi} = -1$) for
any closed loop.
Incidentally in 1983 the Ising Wilson-Majorana links representation
was generalized by
Brower, Giles and Maturana~\cite{PhysRevD.29.704} to Wilson  Fermions in 4d lattice QCD.
On the 4d hypercubic lattice, the  Wilson link  projectors,  $P_\mu = (1 + \gamma_\mu)/2$,  mapped
to  spinor rotations $R_{\mu \nu} = \exp[ i\pi
\sigma_{\mu\nu}/4]$ by $\pi/4$  at each corner in the $\mu$-$\nu$ plane, giving
the appropriate minus sign to  Fermion loops.

Finally, matching the amplitudes of the loop expansion requires corners with weights
\be
\sqrt{ t_i t_j} = \sqrt{\kappa_i \kappa_j} \cos (\theta_{ij}/2)  \quad, \quad \theta_{ij} = \theta_j- \theta_i
\ee
or 
\be
t_1 =  \frac{\kappa_1\cos(\theta_{12}/2)\cos(\theta_{13}/2)}{\cos(\theta_{23}/2)} = \tanh(L_1)  \; .
\ee
It might be interesting to pursue generalizations
of this fermionic loop expansion for general 2d surfaces.   In 2d any triangulated simplicial complex
has a trivalent dual graph leading to a dual non-self-intersecting loop expansion, but
curvature requires the inclusion of
a spin connection following the 
general method in Ref. \cite{Brower:2016vsl}
with manifest gauge invariance with rotation
of the tangent vector at each site.  Similar methods have been used
in the random triangulation solution to 2d string theory~\cite{https://doi.org/10.48550/arxiv.hep-th/9112013}.

\subsection{\label{sec:DiracGeometry} Matching Geometry to Critical CFT on $\mR^2$}

\begin{figure}[h]
    \centering
    \includegraphics[width=0.45\textwidth]{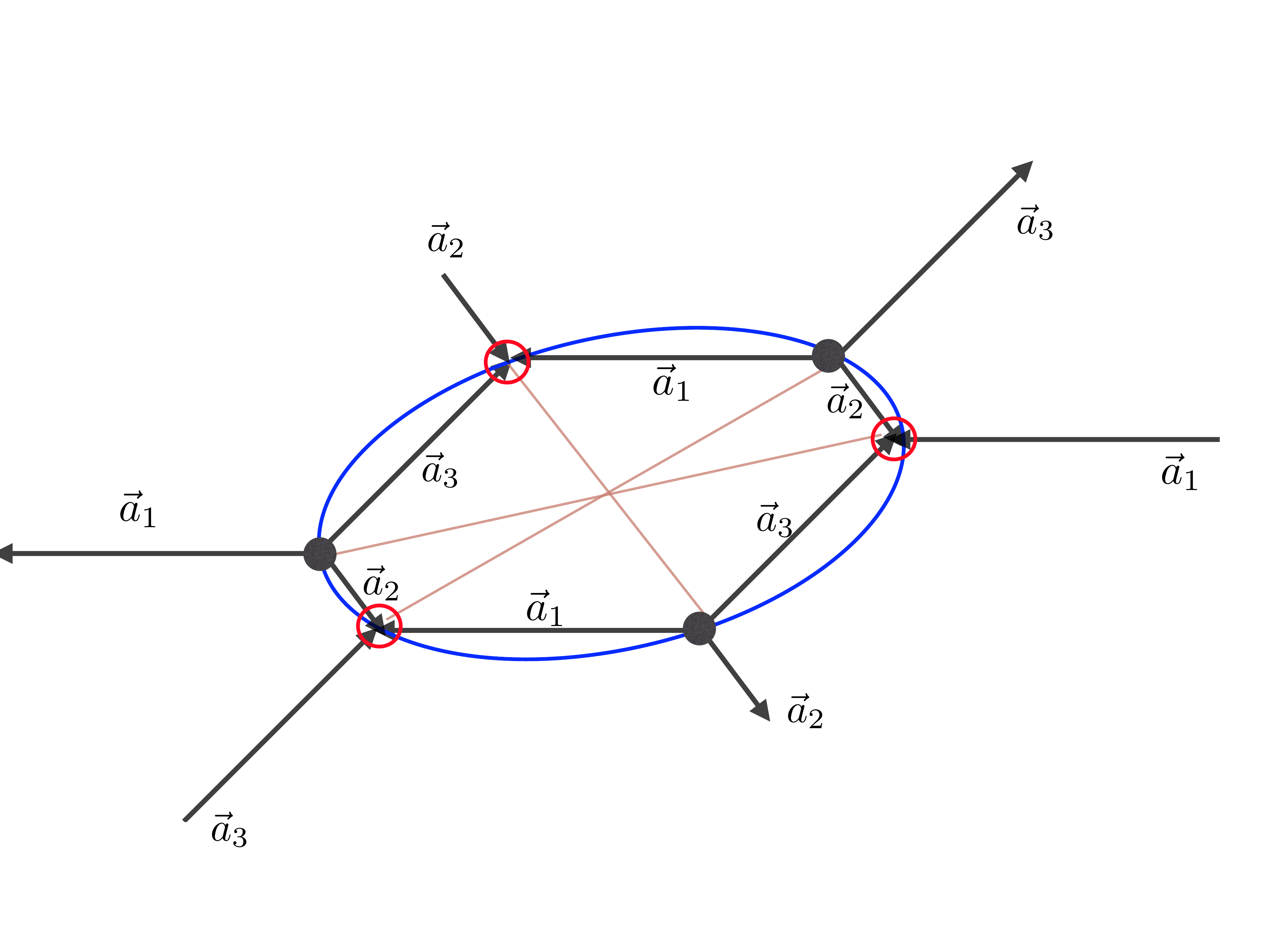}
    \includegraphics[width=0.5\textwidth]{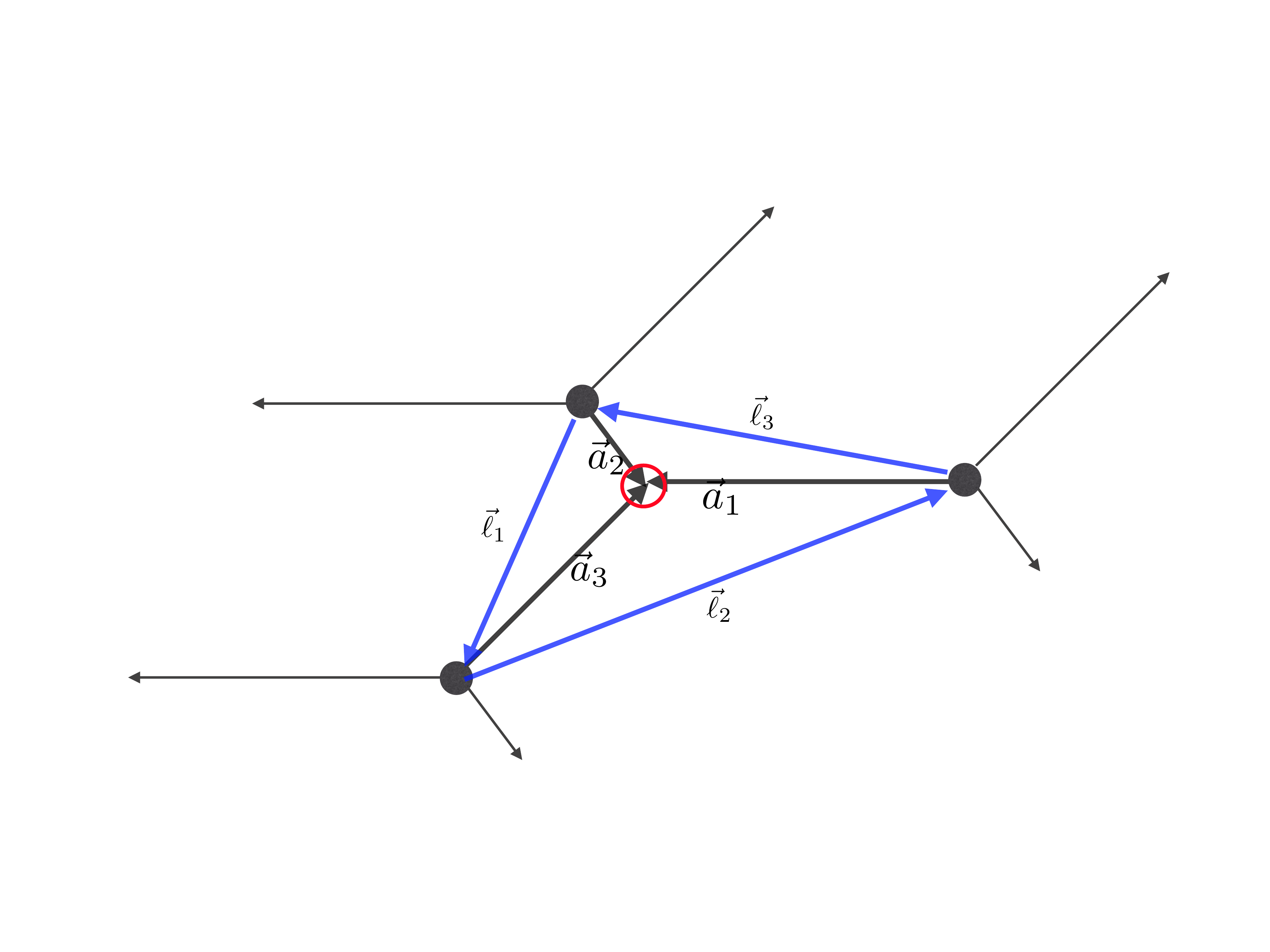}
    \caption{On the left, the edge lengths of parallel links on the hexagonal lattice define a simplicial matrix which on
    the right is applied to the triangular lattice using the star-triangle relation:  $\vec \ell_k = \epsilon^{ijk}  (\vec a_i - \vec a_j )/2$.}
    \label{fig:Ellipse}
\end{figure}

To introduce the full metric information for each hexagonal edge with
coupling constant $L_i$, we assign a 2d vector $\vec a_i  =  a_i \hat e_i  $ leaving
solid black sites as shown on the left in Fig.~\ref{fig:Ellipse}.   We then
use the star triangle  map on the right side of Fig.~\ref{fig:Ellipse}
to  assign edges to the triangular model,
\be
\vec \ell_1 =  \vec a_2 - \vec a_3  \quad \mbox{ and cyclic} 
\ee
or equivalently  $\vec \ell_k = \epsilon_{ijk}  (\vec a_i - \vec a_j )/2$
on the triangular lattice.  Note that this implies correctly 
the triangle condition $\vec \ell_1 + \vec \ell_2 + \vec \ell_2 = 0$.
This is not a one-to-one map.  The geometry for our hexagonal lattice
with opposite parallel edges requires 6 parameters for the three
vectors $ \vec a_i $, while the triangular lattice requires
only 4 parameters.  Our hexagons are special in that they lie on the
boundary of an ellipse which is therefore an example of Pascal's
theorem~\cite{Pascal:1640}, and requires that the intersections
of three pairs of lines for the opposite sides lie on single line.
However in our case with parallel opposite sides the
intersections are all on a line at infinity in projective space.

\begin{figure}[h]
    \centering
    \includegraphics[width=0.6\textwidth]{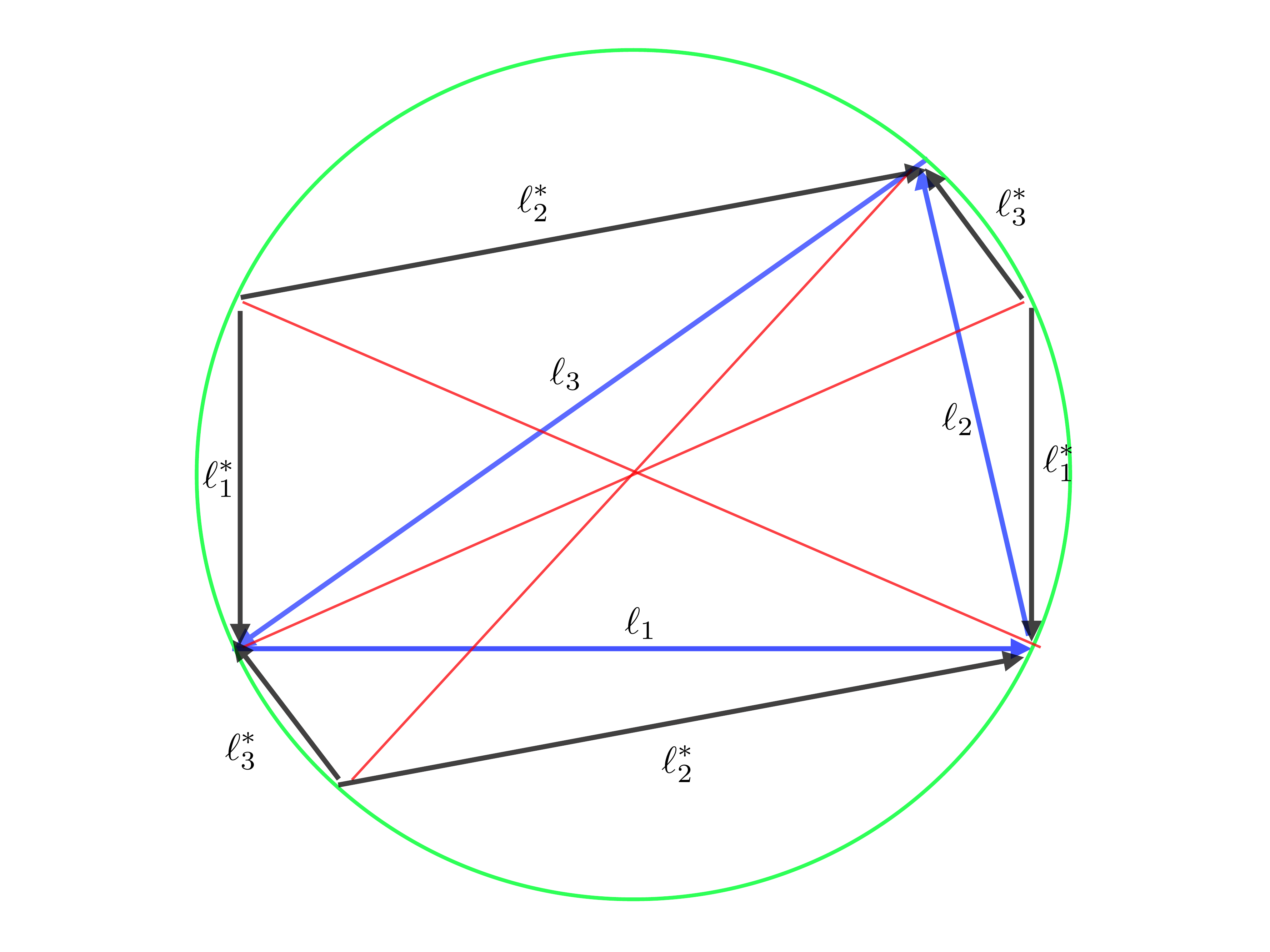}
    \caption{ \label{fig:IDproof}Coincident  triangle (blue)  and hexagon (black) circumscribed by a circle of radius $R$ in green. The
    dual Voronoi vectors $\vec \ell^*_i$ are orthogonal to the triangles vectors $\vec \ell_i$ with magnitude satisfying
    $4 R^2 = \ell^{*2}_i +  \ell^2_i $. }
\end{figure}
    
The two extra degrees of freedom correspond to eccentricity
and the direction of the major axis which exist in the affine transformation of
circles to ellipses.  The affine transformation
of the ellipse to a circle is precisely the metric required to restore
full conformal symmetry  on the critical surface
and the ambiguity of the inverse map is fixed by
requiring the hexagons to be cyclic polygons, i.e. corresponding to the Voronoi dual
of the triangular lattice. This is
illustrated in Fig. \ref{fig:IDproof} showing how the new hexagons and the triangles can both be circumscribed by identical circles.

We now turn to the algebraic proof of this statement, aware
of the fact that there are surely more elegant methods which directly use
projective geometry beyond our current presentation.

Our problem is to find  metric values for $\vec a_i $ that restore rotational symmetry
on the critical surface.  Remarkably, this constrains the hexagonal vectors to
be at the circumcenter dual to the triangular complex, $\vec  a_i  =\vec \ell^*_i $ where  $*$ is the Hodge star of
the discrete exterior calculus (DEC) for the finite element method (FEM) applied to the
free Laplacian in Eq. \ref{eq:BLoperator}.

In anticipation of taking the continuum limit,  we introduce position vectors $x^\mu$
with neighboring sites at  $y^\mu = x^\mu   +  a^\mu_i $. Changing notation to $\kappa_i \rightarrow \kappa_{xy}$ and $\hat e_i \rightarrow \hat e_{xy}$ on  each link $\<x,y\>$, the fermion action  is written
\begin{equation}
  S = \dfrac{1}{2}  \left[ \sum_x \bar \psi_x  \psi_x - \sum_{\<x,y\>}  \kappa_{xy} \bar \psi_x \psi_y  -
  \sum_{\<x,y\>}  \kappa_{xy} \bar \psi_x  \hat e_{xy}  \cdot \vec \sigma\psi_y \right] \; .
 \label{eq:expanded_lattice_action}
\end{equation}
The second term,
\be
 \sum_{\< x,y \>} \kappa_{xy} \bar \psi_x \psi_y=  \sum_{\< x,y \>} \kappa_{xy} \bar \psi_x \psi_x  - \frac{1}{2}\sum_{\< x,y \>}  (\bar \psi_x- \bar \psi_y)(\psi_x - \psi_y) \simeq\frac{1}{2} \sum_i \kappa_i  \sum_x \bar \psi_x \psi_x
\ee
is a mass correction plus a second derivative term which vanishes as the square of the lattice spacing
in the continuum.  Similarly
we may expand the  third term  in (\ref{eq:expanded_lattice_action}) using
\be
\psi_y = \psi_x  + a_{xy} \hat{e}_{xy} \cdot \vec \nabla \psi_i + \mathcal{O}(a^2_i) \; .
\ee
Defining $m_0 = 1 - \sum_i\kappa_i/2$ and noting that $\bar \psi_x \vec{\sigma} \psi_x = 0$ we have
\begin{equation}
    S = \dfrac{1}{2} \left[ \sum_{\<xy\>}  \bar \psi_x  (\hat{e}_{xy} \cdot \vec{\sigma}) ( \kappa_{xy} a_{xy} \hat{e}_{xy} \cdot \vec \nabla )\psi_x  + m_0 \sum_x \bar \psi_x\psi_x \right]  + \mathcal{O}(a_i^2) \; . 
\end{equation}
Near a second order phase we fix the physical  mass $m$ by scaling 
$m_0  = \sqrt{g_x} m$,  where on our regular lattice $ \sqrt{g_x}  =2  A_\triangle$, the
area of the triangle dual to each lattice site on the hexagonal lattice.

Finally, to restore spherical symmetry we must adjust the  metric to satisfy
\be
\sum_i a E_i \partial_i = \sum_i  (\kappa_i\hat e_i \cdot \vec \sigma) ( \vec a_i  \cdot \vec  \nabla ) \sim \vec \sigma \cdot \vec \nabla  \; .
\ee
The solution has a remarkably elegant geometric form by scaling the vectors $a \partial_i = \vec a_{i } \cdot \nabla  = \ell^*_i \hat e_i \cdot \nabla$ aligned along the circumcenter dual edges to  the triangular lattice with edges lengths, $\ell_1,\ell _2, \ell_3$ as shown in Fig. \ref{fig:IDproof}. Now
$\ell^*_i$ is the Hodge star of the edge simplex with length $\ell_i$ and  vierbein $  \vec E_i = \kappa_i \hat e \cdot \vec \sigma =\ell_i  \hat e_i \cdot \vec \sigma $
scaled to the triangular edges. As illustrated in Fig. \ref{fig:IDproof}, corresponding
hexagonal and triangular vectors are orthogonal. Therefore
the vierbein  $ \vec E_i  = -i \sigma_2 \vec \ell_i \cdot \vec \sigma$ is rotated by  90 degrees  relative  to $\vec\ell_i$, in agreement with charge conjugation
of the Wilson-Majorana spinor frame.  The result is that the Hodge duality fixes the vectors to the circumcenter of
the dual triangular simplex.
\be
 \ell^{*\mu}_1 = \epsilon^{\mu \nu}   \ell^\nu_1 ( \ell^2_2 + \ell^2_3 - \ell^2_1)/( 4 A_\triangle)  \quad, \quad  \mbox{1, 2, 3 cyclic}
\ee
This is a 2d example of the  general property of Hodge duality in the FEM discrete exterior calculus
for any simplicial complex and its circumcenter dual. We define a global lattice scale $a$ by
the dual area  of sites in the triangular lattice: $a^2  =  \sqrt{g_x} = 2 A_\triangle $. 

A geometric calculation, easily visualized in the right side of Fig.~\ref{fig:Stokes} gives 
\be
\sum_i  \ell^\mu_i \ell^{*\nu}_i  = 2 A_\triangle \epsilon^{\mu \nu}
\ee
where $2 A_\triangle  =  \vec \ell^*_1 \wedge \vec \ell^*_2 + \vec \ell^*_2 \wedge \vec \ell^*_3  + \vec \ell^*_3 \wedge \vec \ell^*_1$. 
Note the dual area $A_{\hexagon} = 2 A_\triangle = \ell^*_1 \ell_1 +  \ell^*_2 \ell_2 +  \ell^*_3 \ell_3 $.
  
Introducing the  canonical fermion mass $m  = m_0/\sqrt{2 A_\triangle}$ and
rescaling $\bar \psi_x \psi_x \rightarrow \sqrt{2 A_\triangle} \bar \psi_x \psi_x $, the
sum becomes an integral with continuum action
\be
    S \simeq \frac{1}{2} \int d^2 x \bar \psi_x ( \vec{\sigma} \cdot \vec \nabla  + m )\psi_x
\ee
and the critical surface at $m =0$ is given by the constraint
\be
\kappa_i =  \frac{2 \ell_i}{\ell_1 + \ell_2 + \ell_3  } = \frac{ \ell_i}{s } \qquad, \quad  \kappa_1 + \kappa_2 + \kappa_3 = 2
\ee
defining the semi-perimeter $s = (\ell_1 + \ell_2 + \ell_3)/2$. 
The kinetic term fixes the geometry with 
\be
\sinh(2K_i) =1/\sinh (2 L_i)  =  \frac{\ell^*_i}{\ell_i}
\label{eq:crit_couplings} \; .
\ee
This general solution simplifies for the regular triangular lattice  with $\ell^*_i/\ell_i = 1/\sqrt{3}$
giving the critical point  $K_i  =  \ln(3)/4$.  More interesting is 
the limit to rectangular lattice  with  lattices spacing,   $a_x = \ell_1  = \ell^*_2t$,  $a_t =  \ell_2  = \ell^*_1 $, respectively.
This is the $\ell^*_3\rightarrow 0$ limit with action 
 \be
 S = \frac{1}{2} \sum_{x,t} \left[ K_t s_{t,x} s_{t+1,x} +  K_x s_{t,x} s_{t,x+1}   \right] \; .
 \label{eq:rectangular}
\ee
From Eq. \ref{eq:CriticalSurface}, the critical surface is $\sinh(K_t+ K_x) = \cosh(K_t - K_x)$.

From Fig.~\ref{fig:IDproof}, we can visualize the limit  $\ell^*_3 \rightarrow 0$  which forces the diagonal on the triangle to vanish ($K_3   = 0 $) and pairs
of hexagonal spins to be free ($L_3 \rightarrow \infty$) so the
dual is triangular  rectangle is a  hexagonal rectangle, exchanging:  $ a_t \leftrightarrow a_x$.
As mentioned in Sec.~\ref{sec:RadialQ}, these asymmetric lattices  are very useful for radial quantization and finite temperature simulation.

\subsection{\label{sec:Algebra} Algebraic Derivation}
 
The algebraic manipulations to prove Eq.~\ref{eq:crit_couplings} use
standard properties of triangles with sides $\ell_1, \ell_2, \ell_3$.
Heron's formula and the circumradius 
\be
A_\triangle = \sqrt{s(s- \ell_1)(s-\ell_2)(s-\ell_3)}  \quad, \quad R = \frac{\ell_1 \ell_2 \ell_3} { 4 A_\triangle}
\ee
respectively with $s = (\ell_1+\ell_2 + \ell_3 )/2$ as well as the half angle formula for $\theta_{23}$ opposite
the side from $\ell_1$ triangle \footnote{see \href{https://www.cuemath.com/jee/semiperimeter-and-half-angle-formulae-trigonometry/}{https://www.cuemath.com/jee/semiperimeter-and-half-angle-formulae-trigonometry/}}
\be
\cot(\theta_{23}/2) = \sqrt{\frac{s (s-\ell_1 )}{(s-\ell_2 ) (s - \ell_3)}} \; ,
\ee
plus permutations. Putting this together we have
\be
t_1 =  \frac{\ell_1}{ s} \frac{\cos(\theta_{12}/2)\cos(\theta_{13}/2)}{\cos(\theta_{23}/2)}
=   \frac{\ell_1}{  s}  \sqrt{ \frac{ s(s-\ell_3) s(s-\ell_2) } { s(s-\ell_1)^3   }}  
=\frac{ 4  A_\triangle }{ \ell^2_2 + \ell^2_3 - \ell^2_1 + 2 \ell_2 \ell_3} 
=   \frac{\ell_1}{\ell^*_1 + 2 R} \; .
\ee
As a consistency check, using our identity in Eq.~\ref{eq:crit_couplings} we have
\be
t_1 =  \tanh(L_1) =  \frac{ \sinh(2 L_1)}{1 + \cosh(2 L_1) }
=  \frac{\ell_1/\ell^*_1}{1  + \sqrt{1   +( \ell_1/\ell^*_1)^2}} = \frac{\ell_1}{ \ell^*_1 + 2 R} \; .
\ee

Finally, one can show that the zero mass conformal invariant free
fermion, $\sum_i \kappa_i = \sum_i \ell_i/s  = 2$, is equivalent  to  the star-triangle critical surface
\be
t_1 t_2 + t_2 t_3 + t_3 t_1 = \frac{\ell_1 \ell_2}{(\ell^*_1 + 2 R)(\ell^*_2 + 2 R)} + \frac{\ell_2 \ell_3}{(\ell^*_2 + 2 R)(\ell^*_3 + 2 R)} + \frac{\ell_3 \ell_1}{(\ell^*_3 + 2 R)(\ell^*_1 + 2 R)}  = 1
\ee
in Eq. \ref{eq:CriticalSurface} with $t_i = p_i$. 
This is an interesting identity relating a triangular lattice and its Voronoi dual.
 Since, as illustrated in Fig.~\ref{fig:IDproof},
the triangular lengths $\ell_i$ and their circumcenter dual lengths $\ell^*_i$ are
fixed entirely by parallel lines and right angles, it must be
a  consequence of  projective geometry theorems in $\mRP{2}$ which
would be interesting to identify and generalize to  higher dimensions. 

\section{\label{sec:MCR} Monte Carlo Simulations}

In this section we use our formalism to perform Monte Carlo simulations of three interesting examples of the critical Ising model: the finite modular torus $\mT^2$, the infinite Euclidean plane $\mR^2$ (via finite-size scaling), and the cylinder
$\mR \times S^1$ of radial quantization~\cite{Catterall:1989bs}. Of course, any simulation
must start with a finite lattice, so the traditional test of a CFT on
$\mR^2$ uses an $L\times L$ periodic lattice (i.e. a torus) in
the limit where the longest correlation length $\xi$ satisfies $a \ll a \xi \ll a L$ in
units of the lattice spacing $a$. Technically, this should be an extrapolation for the double limit -- first the infinite volume limit $L \rightarrow \infty$ followed
by the approach to the critical surface $\xi \rightarrow \infty$ for a
CFT or massive theory.  However, our affine lattice is also
ideal for comparison with the exact finite volume effects of the
modular invariance of the torus going to the pseudo-critical surface
as we take $L\rightarrow \infty$. Lastly, we consider briefly the advantage
of the approach to the infinite cylinder $\mR \times S^1$ of radial quantization, or
in the limit $L_t \gg L_x$ for finite temperature. All of these have interesting consequences beyond this first test and will be pursued further with an eye to 
higher dimensions. 

For all of our simulations, we perform 2000 thermalization sweeps followed by 50,000 measurement
sweeps, with 20 sweeps between measurements. Each sweep consists of 5
Metropolis updates and 3 Wolff cluster updates \cite{Metropolis1953Equation, wolffCluster1989}. With these
parameters we find that autocorrelations in our measurements are negligible.

\subsection{\label{sec:Modular}The Critical Ising Model on the Modular Torus: $\mT^2$}

The first test of our formalism is to embrace the periodic and finite nature of the lattice and compare our data to the Ising model defined on a 2d torus. Traditionally the Ising model is studied only on square and rectangular lattices, but our formalism allows us to simulate the critical Ising model on a torus with arbitrary modular parameter $\tau$.

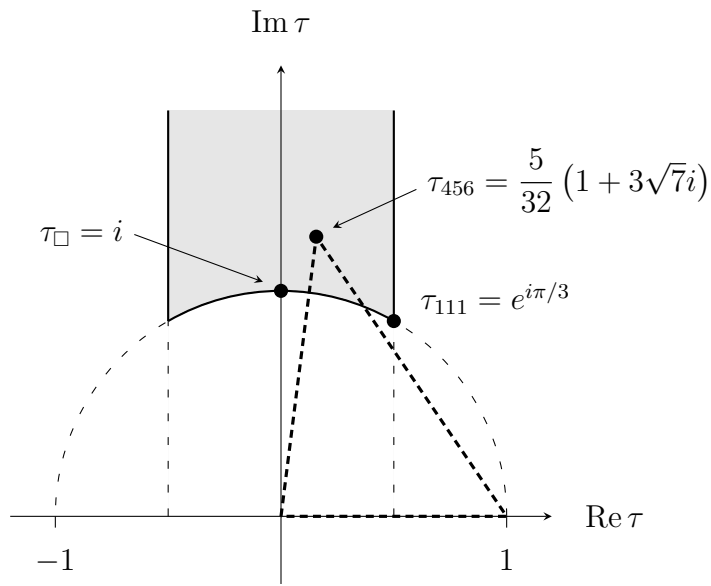
\begin{figure}[h]
    \centering
    \begin{tikzpicture}[>=stealth,scale=3.0]
    
    \draw[->] (-1.2,0) -- (1.2,0);
    \draw[->] (0,-0.3) -- (0,2.0);
    
    \draw[loosely dashed] (1,0) arc (0:180:1);
    \draw[loosely dashed] (0.5,0) -- (0.5,0.866025);
    \draw[loosely dashed] (-0.5,0) -- (-0.5,0.866025);
    \draw[thick] (0.5,0.866025) arc (60:120:1);
    \draw[thick] (0.5,0.866025) -- (0.5,1.8);
    \draw[thick] (-0.5,0.866025) -- (-0.5,1.8);
    
    \node[anchor=west] at (1.3,0) {$\operatorname{Re} \tau$};
    \node[anchor=south] at (0,2.1) {$\operatorname{Im} \tau$};
    \draw (1,0) -- (1,-0.05);
    \draw (-1,0) -- (-1,-0.05);
    \node[anchor=north] at (1,-0.1) {$1$};
    \node[anchor=north] at (-1,-0.1) {$-1$};
    
    \begin{scope}
    \clip (0.5,0)--(0.5,1.8)--(-0.5,1.8)--(-0.5,0)--cycle;
    \fill[fill=black,opacity=0.1] (-0.5,0) rectangle (0.5,2) (0,0) circle[radius=1];
    \end{scope}
    
    \fill (0.5,0.866025) circle[radius=0.03125];
    \fill (0,1) circle[radius=0.03125];
    \fill (0.15625,1.2402) circle[radius=0.03125];
    \node[anchor=south west] at (0.57,0.85) {$\tau_{111} = e^{i \pi / 3}$};
    
    \node[anchor=south east] at (-0.65,1.15) {$\tau_{\Box} = i$};
    \draw[->] (-0.65,1.25) -- (-0.08,1.04);

    \node[anchor=south west] at (0.6,1.3) {$\tau_{456} = \dfrac{5}{32} \left(1 + 3 \sqrt{7} i \right)$};
    \draw[->] (0.6,1.45) -- (0.23,1.28);
    
    \draw[very thick,densely dashed] (0,0) -- (0.15625,1.2402) -- (1,0) -- cycle;
    
    \path(-2,0);
    \path(2,0);
    
    \end{tikzpicture}
    \caption{The fundamental domain of the modular parameter $\tau$.}
    \label{fig:fundamental_domain}
\end{figure}

The modular parameter is a concept familiar to string theorists that is used to parameterize all possible boundary conditions of a torus. Put simply, if the torus is thought of as a tube, the modular parameter is a complex number which parameterizes how the tube is stretched and twisted before its two ends are glued together. The shaded region $\{\tau : |\tau| \ge 1, |\operatorname{Re} \tau| \le 1/2, \operatorname{Im} \tau > 0 \}$ shown in Fig. \ref{fig:fundamental_domain} indicates the fundamental domain of the modular parameter $\tau$. Each value of $\tau$ in this region defines a triangle from which we can construct a unique 2d lattice with periodic boundary conditions and the topology of a torus. We have indicated the locations of the equilateral case $(\tau_{111})$, the square case $(\tau_{\Box})$, and a representative example of a skew triangle $(\tau_{456})$. The heavy dashed lined is the triangle defined by $\tau_{456}$. The notation $\tau_{ijk}$ indicates that the triangle side lengths are proportional to $\{i,j,k\}$.

Without loss of generality, we can sort the triangular lattice lengths so that $\ell_1 \leq \ell_2 \leq \ell_3$. Then the modular parameter in the fundamental domain is
\begin{equation}\label{eq:mod_param}
    |\tau| = \dfrac{\ell_2}{\ell_1}, \qquad \arg(\tau) =  \cos^{-1}(-\hat{e}_1^* \cdot \hat{e}_2^*) \; .
\end{equation}

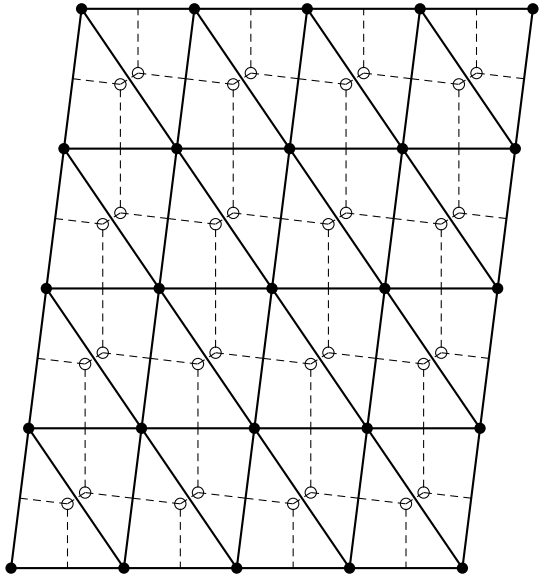
\begin{figure}
    \centering
    \begin{tikzpicture}[scale=1.5]
    
    \foreach \y in {0,1,2,3}
    \foreach \x in {0,1,2,3} {
    \begin{scope}[shift={(\x+\y*0.15625,\y*1.2402)}]
    \draw[thick] (0,0) -- (1,0) -- (0.15625,1.2402) -- cycle;
    \draw[densely dashed] (0.5*0.15625,0.5*1.2402) -- (0.5,0.57) -- (0.5,0);
    \draw[densely dashed] (1+0.5*0.15625,0.5*1.2402) -- (0.5+0.15625,1.2402-0.57) -- (0.5+0.15625,1.2402);
    \draw[densely dashed] (0.5,0.57) -- (0.5+0.15625,1.2402-0.57);
    \draw (0.5,0.57) circle[radius=0.05];
    \draw (0.5+0.15625,1.2402-0.57) circle[radius=0.05];
    \end{scope}
    }
    \draw[thick] (0.625,4*1.2402) -- (4+0.625,4*1.2402) -- (4,0);
    
    \foreach \y in {0,1,2,3,4}
    \foreach \x in {0,1,2,3,4} {
    \begin{scope}[shift={(\x+\y*0.15625,\y*1.2402)}]
    \fill (0,0) circle[radius=0.05];
    \end{scope}
    }

    \end{tikzpicture}
    \caption{The triangular lattice defined by $\tau_{456}$ with $L=4$. The dashed lines indicate the dual hexagonal lattice with sites at the circumcenter of each triangle.}
    \label{fig:dual_lattice}
\end{figure}

The lattice is implemented as a refined parallelogram with triangular cells with the appropriate side lengths as shown in Fig. \ref{fig:dual_lattice}.

The continuum two-point function for the critical Ising model on a torus with modular parameter $\tau$ is known to be~\cite{difrancesco:1987}
\begin{equation}\label{eq:torus_exact}
    \langle \sigma(0) \sigma(z) \rangle = \left| \dfrac{\vartheta_1'(0 | \tau)}{\vartheta_1(z | \tau)} \right|^{1/4} \dfrac{\sum_{\nu=1}^4 |\vartheta_{\nu}(z/2|\tau)|}{\sum_{\nu=2}^4 |\vartheta_{\nu}(0|\tau)|}
\end{equation}
where $\vartheta_{\nu}(z|\tau)$ are the Jacobi theta functions and $z = x + iy$ is the separation vector in complex coordinates. We will use this formula to test our result for the critical Ising couplings on a lattice with an arbitrary modular parameter.

\begin{figure}[h]
    \centering
    \includegraphics{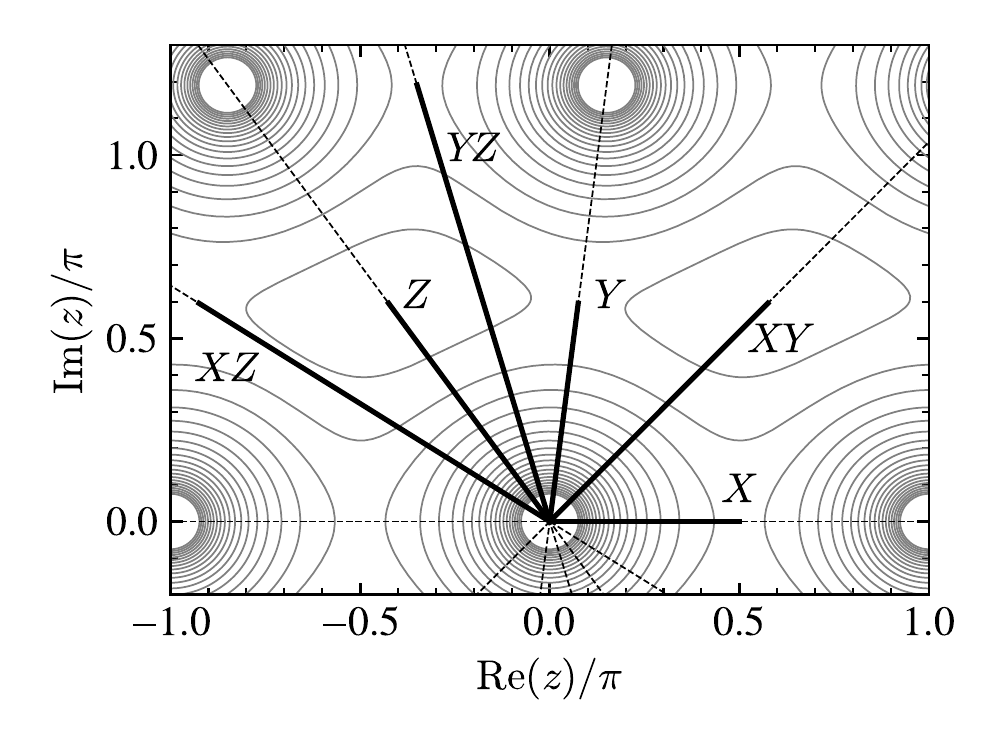}
    \caption{The continuum two-point function in the $z$-plane with modular parameter $\tau_{456}$, highlighting six axes along which we can easily measure the two-point function on our lattice. }
    \label{fig:torus_2pt_contour}
\end{figure}

In Fig. \ref{fig:torus_2pt_contour} we show a contour plot of the continuum 2d Ising spin-spin two-point function on a torus with modular parameter given by Eq.~\ref{eq:mod_param} for a triangle with side lengths $\ell_i \propto \{ 4,5,6 \}$ and coupling coefficients $\sinh 2 K_i = \ell_i^* / \ell_i$, which gives $K_i \simeq \{ 0.48648, 0.31824, 0.062829 \}$. On a triangular lattice, it is convenient to measure the two-point function along the six axes shown. It is sufficient to measure only the bold part of each axis due to the periodicity of the torus.

\begin{figure}
    \centering
    \includegraphics{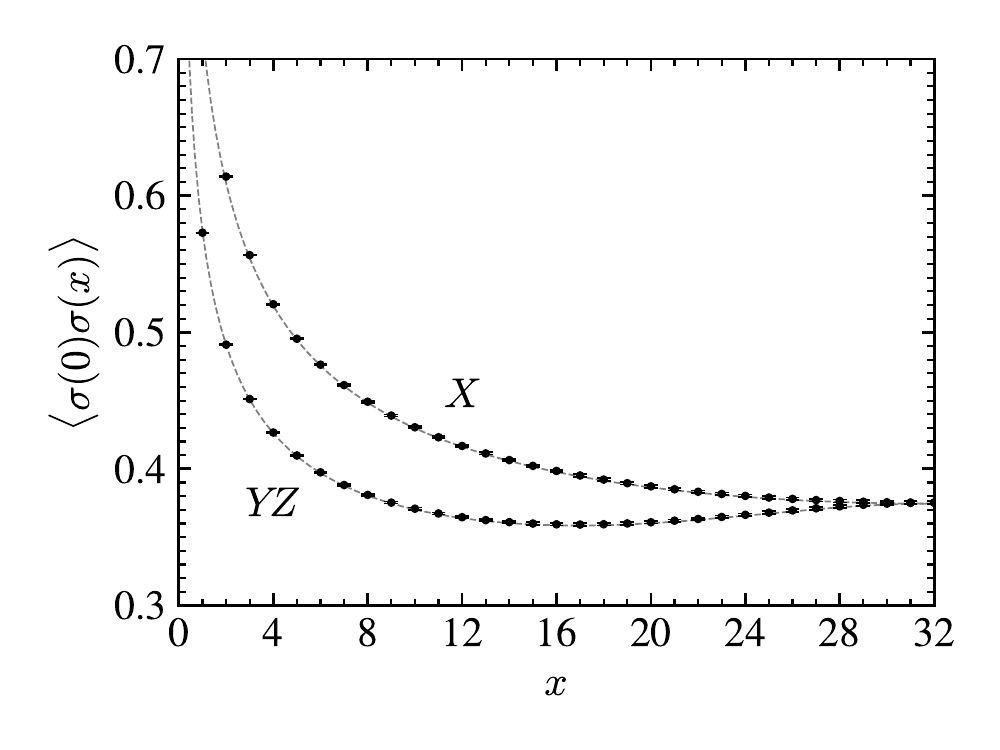}
    \caption{Two-point correlation function measured along the $X$ and $YZ$ axes shown in Fig. \ref{fig:torus_2pt_contour} for a triangular lattice with $\ell_i \propto \{ 4,5,6 \}$ and $L=64$. The horizontal axis is the distance measured in lattice steps. The gray lines are the exact correlation function from Eq. \ref{eq:torus_exact} and the black points are lattice data.}
    \label{fig:ising_flat_crit}
\end{figure}

We perform a simultaneous fit to Eq. \ref{eq:torus_exact} using lattice data for all six of these axes. The only fit parameter is an overall normalization factor. We can see in Fig. \ref{fig:ising_flat_crit} that the lattice data is in excellent agreement with the continuum result.

\FloatBarrier
\subsection{\label{sec:MCR2} The Critical Ising Model on $\mR^2$: Finite-Size Scaling}

In this section, we use the same lattice construction as in Sec.~\ref{sec:Modular}, but now we extract information about the continuum theory in an infinite plane via finite-size scaling. A similar analysis was done for an equilateral triangular lattice in \cite{Zhi-HuanLuo2009Cbot}. Our formalism allows us to construct the lattice from triangles with any side lengths. Here, we again use triangles with $\ell_i \propto \{ 4,5,6 \}$ because it is sufficiently different from an equilateral, isosceles, or right triangle and will show a clear difference between measurements along different triangle axes.

\begin{figure}[h]
    \centering
    \includegraphics{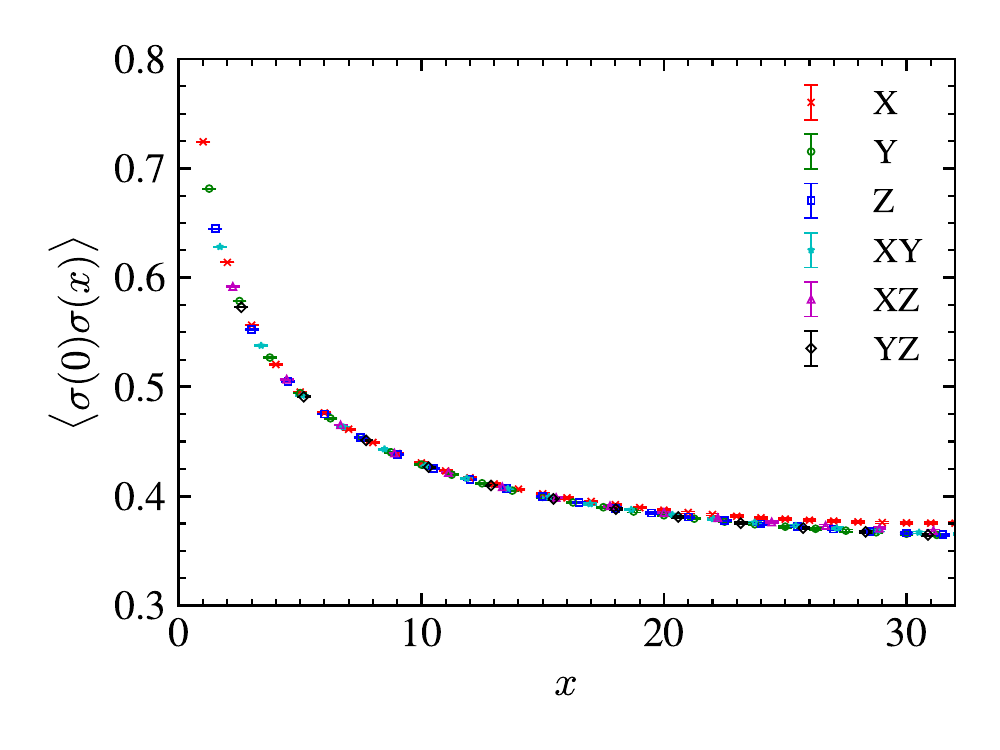}
    \caption{Spin-spin correlation function measured along each of the axes shown in Fig. \ref{fig:torus_2pt_contour} with distances scaled appropriately based on the lattice lengths $\ell_i$. Shown here for $\ell_i \propto \{ 4,5,6 \}$.}
    \label{fig:ising_r2_crit}
\end{figure}

Again using critical couplings $\sinh 2 K_i = \ell^*_i / \ell_i$, we measure the spin-spin correlation
function along the six axes shown in
Fig. \ref{fig:torus_2pt_contour}. After scaling the step length for
each axis appropriately, the correlation functions collapse onto a single curve
at small separation, as shown in Fig. \ref{fig:ising_r2_crit}. At
large separation, wraparound effects cause the correlation function behavior to
depend slightly on which axis is being measured.

\begin{figure}[h]
    \centering
    \includegraphics{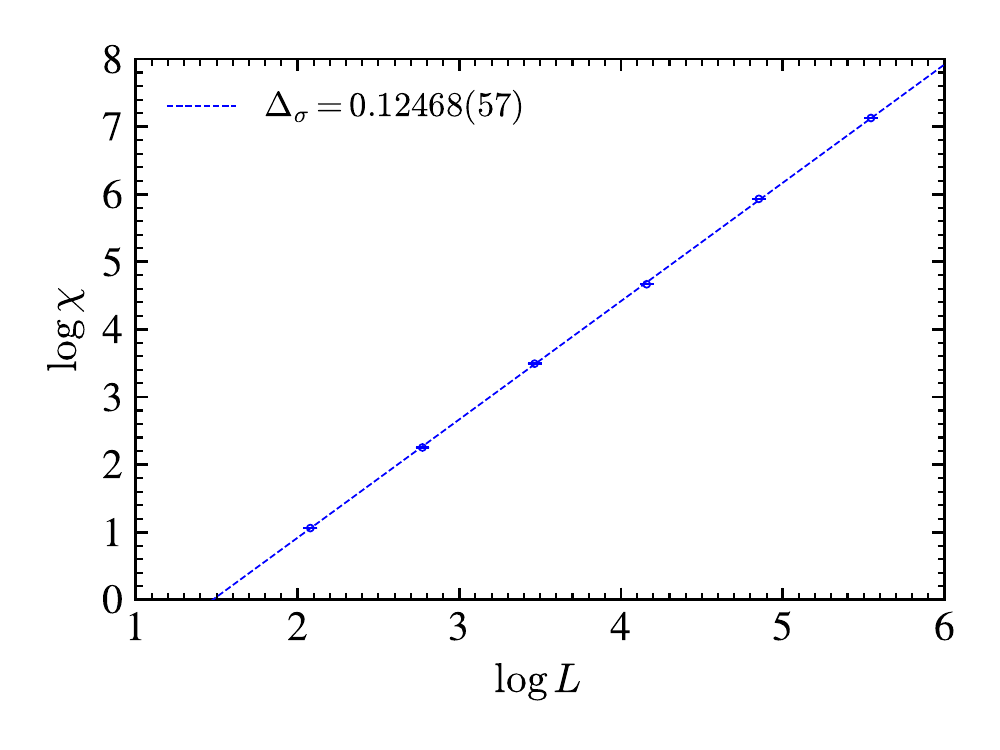}
    \caption{Finite-size dependence of magnetic susceptibility on a lattice with $\ell_i \propto \{ 4,5,6 \}$.}
    \label{fig:ising_chi_crit}
\end{figure}

We perform a finite-size scaling analysis \cite{FisherMichaelE.1972STfF, LandauPhysRevB.14.255, Binder:1981sa} to extract the scaling exponent of the first $\mathbb{Z}_2$-odd primary operator, $\Delta_\sigma$. We measure the magnetic susceptibility $\chi = \langle m^2 \rangle - \langle |m| \rangle^2$ where $m = \sum_i \sigma_i$ is the magnetization. On a finite lattice with characteristic size $L$, the magnetic susceptibility should scale as $\chi(L) \propto L^{2-2 \Delta_{\sigma}}$. Fitting measurements on $L \times L$ triangular lattices for $L = 8$ up to $L = 256$, we find $\Delta_{\sigma} = 0.12468(57)$, in excellent agreement with the exact continuum value $\Delta_{\sigma} = 1/8$. Our fit is shown in Fig. \ref{fig:ising_chi_crit}.

 \FloatBarrier
\subsection{\label{sec:RadialQ} Determination of Scaling via Radial Quantization}

Here we present the lattice radial quantization method for measuring the correlation functions of conformal theories. In radial quantization, we perform a Weyl transformation
\begin{equation}
    ds^2_{\text{flat}} = r^2[(d \log r)^2 + d \Omega_{d-1}^2] \to ds^2_{\text{cyl.}} = dt^2 + d \Omega_{d-1}^2
\end{equation}
which takes the flat Euclidean manifold to a cylinder with spherical
cross-section, $\mathbb{R}^d \to \mathbb{R} \times S^{d-1}$. The
scale is fixed relative the radius of the sphere, which we have
set to 1 by convention.  In practice, this is very difficult in
$d>2$ because there are no trivial lattice representations of
$S^{d-1}$. In 2 dimensions however, we can simply reinterpret a
typical periodic lattice as the manifold $\mathbb{R} \times S^1$ of
radial quantization. The angular direction is inherently periodic,
and therefore does not introduce unwanted wraparound effects. The
periodic boundary conditions in the lattice radial direction still
result in finite-size effects, but because of the Weyl
transformation, equally-spaced lattice steps in the radial
coordinate $t = \log r$ are actually exponentially spaced in the
radial separation distance $r$. This means that conformal
correlation functions, which have power-law scaling behavior, decay
exponentially in lattice steps. Therefore wraparound effects are exponentially suppressed as the radial lattice size becomes large. This
allows us to extract scaling dimensions for the operators of
conformal theories using methods traditionally used to extract
energies from operators in gapped theories.

The formalism presented in the previous sections allows us to simulate the radially quantized critical 2d Ising model with unequal lattice spacing in the radial and angular directions. For the results in this section we use a rectangular lattice
\be
    S = \frac{1}{2} \sum_{x,t} \left[ K_t s_{t,x} s_{t+1,x} +  K_x s_{t,x} s_{t,x+1}   \right]
\ee
described in Eq.~\ref{eq:rectangular}.

In the continuum, the exact two point function for the lowest $\mathbb{Z}_2$-odd primary operator $\sigma(t,\theta)$ on the infinite cylinder  is
\be
    G(t,\theta)= \< \sigma(t_2,\theta_2) \sigma(t_1,\theta_1)\> \propto \frac{1} {(\cosh t  - \cos\theta)^{\Delta_\sigma}}
\ee
with $t =t_1 - t_2 $ and $\theta = \theta_1 -\theta_2 $ and $\Delta_\sigma =1/8$. On the lattice, we measure the Fourier coefficients of the two-point function, which we define as
\begin{equation}
    C_m(t) = \sum_x \sigma(0) \sigma(t,x) \cos(2 \pi m x / L_x)
\end{equation}
where $m$ is an integer and $(x,t)$ are the angular and radial separation in lattice units. We define a dimensionless ``speed of light'', $c = 2 \pi a_t / L_x a_x$. Ignoring finite size effects, for large $t$ we expect a result with the form
\begin{equation}\label{eq:rad_corr}
    C_m(t) \propto e^{-c \Delta_{\sigma}^{(m)} t}
\end{equation}
where $\Delta_{\sigma}^{(0)} = \Delta_{\sigma}$ is the scaling exponent of the conformal primary operator and we expect the scaling exponents of the conformal descendant operators to be integer spaced in the continuum limit, i.e. $\Delta_{\sigma}^{(m)} \to \Delta_{\sigma} + m$ as $L \to \infty$. The $m=0$ coefficient is periodic in $t$ so we improve our fit by fitting to
\begin{equation}\label{eq:rad_corr_0}
    C_0(t) \propto e^{-c\Delta_{\sigma} t} + e^{-c\Delta_{\sigma} (L_t - t)} \; .
\end{equation}
We use an asymmetrical lattice with $L_t/L_x = 8$ and $a_t / a_x = 3/2$. Because we are using a rectangular lattice we can relate the lattice spacings to triangle side lengths via $a_x = \ell_x = \ell^*_t$ and $a_t = \ell_t = \ell^*_x$. The Ising coupling constants in the radial and angular directions are then given by $\sinh 2 K_t = 2/3$ and $\sinh 2 K_x = 3/2$, respectively. We note that the coupling constants for the diagonal links are zero and can therefore be neglected. On this lattice, finite-size effects from wraparound in the radial direction lead to corrections of order $\mathcal{O}(\exp(-c \Delta_{\sigma} L_t) \simeq 10^{-5})$.

\begin{figure}[h]
    \centering
    \includegraphics[width=0.49\textwidth]{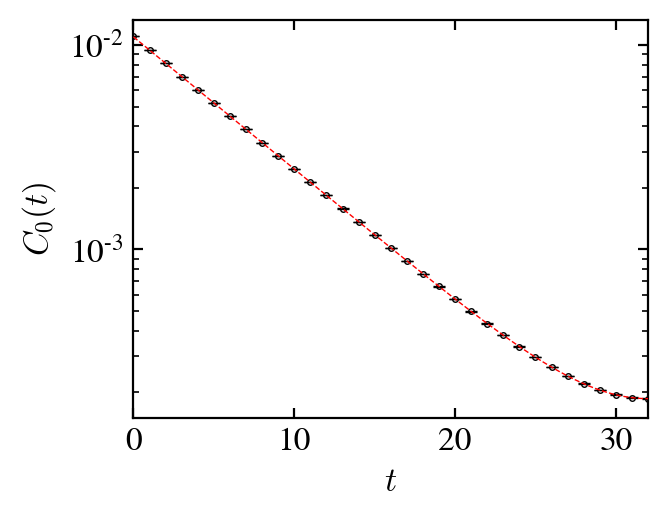}
    \includegraphics[width=0.49\textwidth]{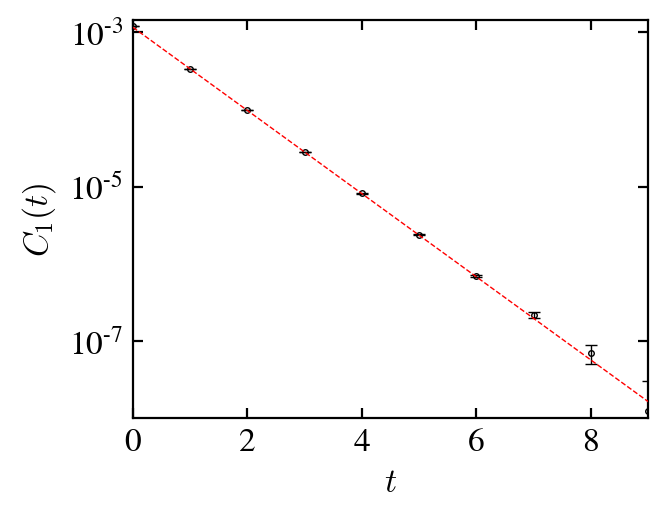}
    \caption{Examples of fits to Eq. \ref{eq:rad_corr} and \ref{eq:rad_corr_0} to extract the scaling behavior of the spin-spin two-point function in radial quantization. These plots are from an $8 \times 64$ rectangular lattice with $a_t / a_x = 3/2$.}
    \label{fig:ising_radial}
\end{figure}

\begin{figure}[h]
    \centering
    \includegraphics[width=0.49\textwidth]{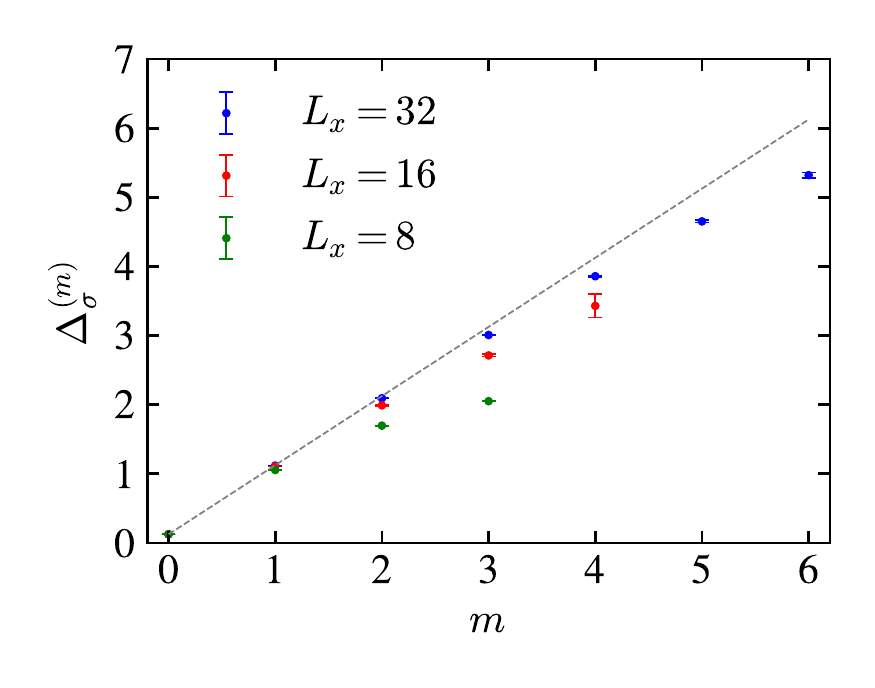}
 \includegraphics[width=0.49\textwidth]{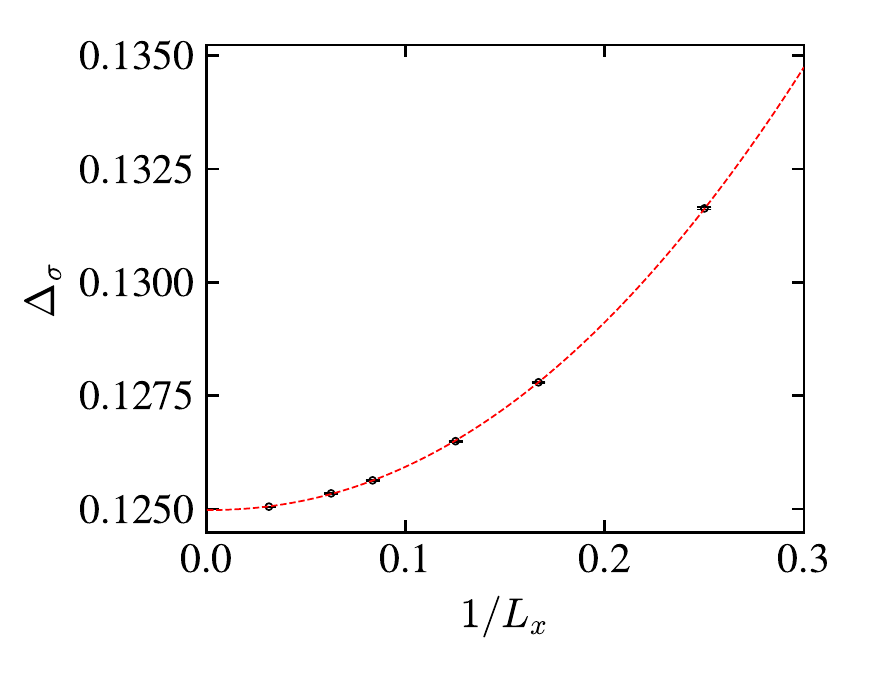}
    \caption{On the left the approximate integer spacing of descendant
      scaling exponents from rectangular lattices with $L_t = 8 L_x$
      and $a_t / a_x = 3/2$. The dashed line shows the continuum
      expectation of exact integer spacing from the primary,
      i.e. $\Delta_{\sigma}^{(m)} \to \Delta_{\sigma} + m$. On the right the continuum limit extrapolation of the scaling operator $\Delta_{\sigma}$. The red line is a fit to Eq. \ref{eq:delta_cont}.}
    \label{fig:rad_int_cont}
\end{figure}

Some example fits are shown in Fig. \ref{fig:ising_radial}. We use a
Bayesian model-averaging procedure \cite{Jay_2021} to average over
choices of the minimum $t$ value to include in each fit. We obtain our
estimate for $\Delta_{\sigma}$ from the $m=0$ fit only. As shown in
Fig. \ref{fig:rad_int_cont} on the left, we find that the descendant scaling exponents $\Delta_{\sigma}^{(m)}$ approach integer spacing as the lattice spacing goes to zero.

Finally, we measure $\Delta_{\sigma}(L_x)$ on lattices with increasing $L_x$ to extrapolate to the continuum limit. We parameterize the finite-volume effects by fitting the finite lattice exponents to the form
\begin{equation}
\label{eq:delta_cont}
    \Delta_{\sigma}(L_x) = \Delta_{\sigma}(\infty)  + b /L_x^{\gamma} 
\end{equation}which gives an infinite-volume scaling exponent of
$\Delta_{\sigma}(\infty) = 0.1249781(62)$ as shown in
Fig. \ref{fig:rad_int_cont} on the right. For the other two fit parameters, we obtain $b = 0.1245(22)$ and $\gamma = 2.114(11)$.

\FloatBarrier

The asymmetric rectangular lattice has advantages which we will
exploit in future efforts for $\mR \times S^d$. Using finer lattice spacing in the radial direction
$a_t \ll a_x$ may improve the extractions of conformal data for 2-point and 4-point functions. It is also interesting to study the
finite temperature regime with fixed $\beta = a_t L_t \ll 1$. The expressions for the couplings $K_x, K_t$ correspond to the Karsch
coefficients \cite{KARSCH1982285} on an asymmetric lattice used in the
extensive study of finite temperature studies for Lattice QCD. With
the analytical form derived here, derivatives with respect to temperature
$k T = 1/\beta$ can be performed, leading to an efficient lattice application of
the Ising  thermodynamics formulated  in Ref.~\cite{Delacr_taz_2022}.  Finally, to simulate the limit
$a_t/a_x \to 0$, there are efficient continuous time worm-type Wolff
cluster algorithms \cite{PhysRevE.66.066110,PhysRevE.69.066129}. As
pointed out by Deng and Bl\"ote\cite{deng2003conformal} this is
equivalent simulating the partition function $\operatorname{Tr}[ \exp( -t H_{\text{QM}} )]$
for the quantum spin Ising Hamiltonian,
\be
H_{\text{QM}} = - \sum_{x,t} ( \sigma^1_x + g^2 \sigma^3_x\sigma^3_{x+1})
\ee
which has critical coupling is $g_c = 1$. We expect to
pursue this in the future with high-precision calculations.

\section{\label{sec:Conclusion} Conclusion}

We have shown that the 2d Ising model on a general uniform triangular grid maps
to the $c= 1/2$ minimal model on $\mR^2$ if the lattice triangulation is
given the appropriate affine transformation of the equilateral lattice.
This we view as a simple and fortunately analytically soluble example
of geometry emerging from a strong coupling quantum field theory at a
second order critical surface.  This affine lattice allowed us to
simulate the toroidal geometry and take the continuum limit consistent
with exact modular invariance.

We have used this example to suggest that there is a natural bridge
between the projective geometry formalism for the CFT and the affine
properties of finite elements on the simplicial complex and its Voronoi
circumcenter dual.  We note that the finite element method (FEM)
provides a general solution for {\em free conformal theories}, defining the
map from simplicial complex to continuum CFT on any
smooth Riemann manifold in any dimension.  While there is much more to
consider in this geometric framework, it does suggest possible
extensions to other strongly coupled CFTs.

First, in 2d flat space there are many exactly soluble theories
which might provide additional examples for this map
from lattice action parameters to continuum CFTs.
Even beyond soluble models, numerical methods
in higher dimensions could be performed to
see if similar maps exist in a finite local parameter space, with the hope that the language of projective geometry will suggest
an \textit{ansatz} to test.

Second, based on our observation that
affine transformations provide a general approach to mapping regular lattices in
flat space locally to the tangent plane on the curved manifold
to $\mathcal{O}(a^2)$ in the lattice spacing $a$, it appears
with preliminary numerical investigations
that demanding a smooth  $\mathcal{O}(a^2)$ 
change in the affine connection between local tangent planes
may enable the local determination of a lattice
action in  a finite  parameter space to reach
the  continuum quantum field theory for the target geometry on the
curved manifold. This would be a start at the quantum generalization of
the QFE program to  strong coupling CFTs. Initial
efforts to apply this affine lattice  framework to the
tangent plane for radial quantization on $\mR \times S^1$
and  $\mR \times S^2$ appear promising. 

We acknowledge the challenges in both of these directions but with
deeper insight into the connection between conformal
geometry and the simplicial lattice calculus we believe some further
progress appears likely. 

\section*{Acknowledgements}
We thank Cameron Cogburn, George Fleming, Ami Katz, Curtis Peterson and  Chung-I Tan for helpful discussions.
This work was supported by the U.S. Department of Energy (DOE) under Award No.~DE-SC0019139 and Award No.~DE-SC0015845. The research reported in this work made use of computing and long-term storage facilities of the USQCD Collaboration, which are funded by the Office of Science of the U.S. Department of Energy.

\bibliography{main}

\end{document}